\documentclass[fleqn,usenatbib]{mnras}

\usepackage{newtxtext,newtxmath}

\usepackage[T1]{fontenc}

\DeclareRobustCommand{\VAN}[3]{#2}
\let\VANthebibliography\thebibliography
\def\thebibliography{\DeclareRobustCommand{\VAN}[3]{##3}\VANthebibliography}

\usepackage{xcolor}
\usepackage{longtable}
\usepackage{supertabular}
\usepackage{rotating}
\usepackage{pdflscape}
\usepackage{geometry}
\usepackage{graphicx}	
\usepackage{amsmath}	
\usepackage{indentfirst}
\usepackage{longtable}  
\usepackage{placeins}   
\usepackage{color}     
\usepackage[symbol]{footmisc}
\usepackage{fp}
\usepackage{multirow}
\usepackage{footnote}
\usepackage{natbib}
\usepackage{hyperref}
\hypersetup{
    colorlinks=true,
    linkcolor=blue,
    filecolor=magenta,
    urlcolor=blue,
}

\title[Tidal interactions in a star-brown dwarf system]{The first evidence of tidally induced activity in a brown dwarf-M dwarf pair: A Chandra study of the NLTT 41135/41136 system}

\author[N. Ili\'c et al.]{
Nikoleta Ili\'c$^{1,2}$\thanks{E-mail: nilic@aip.de},
Katja Poppenhaeger$^{1,2}$,
Desmond Dsouza$^{1,2}$,
Scott J.~Wolk$^{3}$,
Marcel A.~Ag\"{u}eros$^{4,5}$,
\newauthor Beate Stelzer$^{6}$
\\
$^{1}$Leibniz Institute for Astrophysics Potsdam (AIP), An der Sternwarte 16, 14482 Potsdam, Germany\\
$^{2}$Universit\"at Potsdam,  Institut f\"ur Physik und Astronomie,  Karl-Liebknecht-Stra\ss e 24/25, 14476 Potsdam, Germany\\
$^{3}$ Harvard–Smithsonian Center for Astrophysics, 60 Garden Street, Cambridge, MA 0213, USA\\
$^{4}$ Department of Astronomy, Columbia University, 550 West 120th Street, New York, NY 10027, USA\\
$^{5}$ Laboratoire d’astrophysique de Bordeaux, Univ. Bordeaux, CNRS, B18N, Allée Geoffroy Saint-Hilaire, 33615 Pessac, France\\
$^{6}$ Institut f\"ur Astronomie \& Astrophysik, Eberhard-Karls-Universit\"at T\"ubingen, Sand 1, 72076 T\"ubingen, Germany\\
}

\date{Accepted XXX. Received YYY; in original form ZZZ}

\pubyear{2022}

\begin{document}
\label{firstpage}
\pagerange{\pageref{firstpage}--\pageref{lastpage}}
\maketitle

\begin{abstract}
The magnetic activity of low-mass stars changes as they age. The primary process decreasing the stellar activity level is the angular momentum loss via magnetized stellar wind. However, processes like tidal interactions between stars and their close companions may slow down the braking effect and the subsequent decrease of the activity level. Until now, the tidal impact of substellar objects like brown dwarfs on the evolution of their central stars has not been quantified. Here, we analyse the X-ray properties of NLTT~41135, an M dwarf tightly orbited by a brown dwarf, to determine the impact of tidal interactions between them. We find that NLTT~41135 is more than an order of magnitude brighter in the X-ray regime than its stellar companion NLTT~41136, also an M dwarf star, with whom it forms a wide binary system. To characterize the typical intrinsic activity scatter between coeval M dwarf stars, we analyse a control sample of 25 M dwarf wide binary systems, observed with XMM-Newton and Chandra telescopes and the eROSITA instrument onboard the Spectrum R\"ontgen Gamma satellite. The activity difference in the NLTT~41135/41136 system is a $3.44 \sigma$ outlier compared to the intrinsic activity scatter of the control systems. Therefore, the most convincing explanation for the observed activity discrepancy is tidal interactions between the M dwarf and its brown dwarf. This shows that tidal interactions between a star and a substellar companion can moderately alter the expected angular-momentum evolution of the star, making standard observational proxies for its age, such as X-ray emission, unreliable.
\end{abstract}




\section{Introduction}

\noindent Stellar magnetic activity - the collective name for magnetic phenomena of low-mass, main sequence stars\footnote{Mainly stars with an outer convective layer and masses below 1.2 $\mathbf{M\odot}$.} such as coronal X-ray emission, star spots, flares, etc., - is ultimately driven by stellar rotation through the dynamo process. In general, the rotational evolution of a star is determined by its initial spin, its pre-main-sequence contraction rate, and the efficiency of magnetic wind. 

The magnetized stellar wind is particularly important, as it carries away angular momentum from the star. This process, called magnetic braking, slows down the rotation rate of cool stars over timescales of Gyr and weakens the aforementioned magnetic phenomena \citep{Kraft1967,WeberDavis1967,Mestel1968,Skumanich1972,Belcher1976}.

However, if a star has a close-in companion, tidal interactions may alter the stellar rotation and activity evolution described by the spin-down paradigm. This is well-studied for close stellar binaries where the stellar spins are tidally synchronized with the orbital period of the binary. There, the angular momentum loss through stellar winds is replenished from the large angular momentum reservoir of the orbital motion of the two stars \citep{Zahn1977,Hut1981,Terquem1998}. Consequently, close binaries are commonly observed to be highly active even when their ages reach Gyr \citep{Yakut2009}.

Whether substellar close-in companions are able to alter the rotational evolution of a star has been a long-standing question. The so-called Hot Jupiters - close-in massive exoplanets - are the usual suspects in this regard, and many studies have employed different methods to find indications that stars hosting these planets are more active and have a higher spin rate than similar planet-free stars \citep{Cuntz2000,Kashyap2008,Pont2009,Scharf2010,Miller2015,Maxted2015}.

However, obtaining observational confirmation of the increased spin and activity of Hot Jupiter hosts is hard. The main obstacles are the intrinsic variability of stellar magnetic activity \citep{Baliunas1995,Judge2003,Robrade2012,Reinhold2017}, the detection biases against finding exoplanets around magnetically active stars \citep{Poppenhaeger2010,PoppenhaegerSchmitt2011}, as well as a fundamental difficulty in determining ages of single field stars \citep{Weiss1998,Lachaume1999,Pont2004,Valle2015}.

One approach to overcome the problem of the stellar ages has been introduced by \cite{Poppenhaeger2014}, who used wide binary star systems as a coeval laboratory in which one can test if the activity of the star hosting a potentially tidally interacting body is elevated compared to the coeval companion star. We would expect two stars of very similar masses and with the same age to display similar levels of activity. A clear over-activity of the planet-hosting star would provide an indication that tidal prevention of stellar spin-down is at work. By applying a similar method to a large sample, \cite{Ilic2022} found that Hot Jupiter-hosting stars can have their activity level elevated by a factor of $\approx 3$ in the X-ray regime when compared to their coeval companion.

Although this difference is significant and follows a clear trend between stellar activity and expected tidal interaction strength, it is also known that the usual stellar variability throughout an activity cycle can be of a similar order of magnitude\footnote{For the system HD~189733, which according to \cite{Ilic2022} shows the highest activity discrepancy between the coeval stars, \cite{Pillitteri2022} have found no indication for the existence of an activity cycle or significant variability of the planet host on the timescales considered in their work.} \citep{Favata2008,DeWarf2010,Ayres2014,Orlando2017}. Applying the method established by \cite{Poppenhaeger2014} on a system where one late-type star is orbited by a high-mass sub-stellar companion should yield a more significant result in favor of the tidal-interaction hypothesis. We, therefore, explore a mass regime above Hot Jupiters, but below stellar binaries: brown dwarfs orbiting low-mass stars. Brown dwarfs are typically heavier by a factor of $\approx 30$ compared to Jupiters and should have stronger and therefore more definitively measurable tidal effects on their host stars.

For this purpose, we analyse a binary system consisting of two M~dwarf stars, the primary NLTT~41136 and secondary NLTT~41135, where the secondary is orbited by a brown dwarf in close orbit. To determine the significance of tidal interactions, we compare the measured activity difference in this system to the activity differences in a control sample of wide binary systems that have stars of similar spectral types. The stellar activity indicator we choose is the X-ray surface flux since it is the best tracer of the average coronal temperature among the usual activity indicators in the X-ray regime and is, therefore, a good representation of the overall coronal activity of a star \citep{Johnstone2015}.

In Section 2, the {\it Chandra} X-ray observation of the NLTT~41135/41136 system is analysed: we estimate the average coronal temperature, the stellar radius, and calculate the X-ray surface flux. The control sample is introduced in Section 3, along with the results of the analysis of the activity difference in these systems and NLTT~41135/41136. Section 4 opens the discussion on the activity difference distribution of the control sample and how the activity difference in NLTT~41135/41136 compares to these findings. In Section 5, we conclude with a discussion of the significance of the tidal interactions between the M dwarf and the orbiting brown dwarf in NLTT~41135.

\section{Observations and Analysis of NLTT~41135/41136}
\label{sec:analysis}

\subsection{The system}

\noindent NLTT~41135/41136 is a low-mass stellar binary system. It contains two M dwarf stars - NLTT~41136 with spectral type M4V and mass 0.21~$\mathrm M\odot$, and NLTT~41135 with spectral type M5V and mass 0.16~$\mathrm M\odot$ \citep{Irwin2010}. The proper motions of the two stars imply that they form a gravitationally bound and therefore coeval system \citep{Mugrauer2019}, with an angular separation of $2.3\arcsec$ ($\approx 80$ AU). The secondary, NLTT~41135, is orbited by a transiting brown dwarf with a mass of 31-34~$\mathrm{M_{Jup}}$ and an orbital period of 2.9 days \citep{Irwin2010}. By deriving the galactic velocity of the system, \cite{Irwin2010} found that it belongs to the old Galactic disk population, suggesting that this system is older than a few Gyr. The spectral type of the brown dwarf is undetermined; however, its mass and the age of the system indicate a spectral type between T6 and T8 \citep{Pecaut2013,Filippazzo2015}\footnote{We used the mass and age estimate to find the most probable temperature of the brown dwarf using the sample analysed by \cite{Filippazzo2015}; with the temperature estimate, we used the \href{https://www.pas.rochester.edu/~emamajek/EEM_dwarf_UBVIJHK_colors_Teff.txt}{main-sequence parameter table} defined by \cite{Pecaut2013}, which includes parameters of brown dwarfs, to estimate the spectral type.}.

The spectroscopic observation of the system yielded the $\mathrm{H\alpha}$ line in emission in the spectrum of NLTT~41135, while in the spectrum of NLTT~41136, there is a hint of absorption at this wavelength \citep{Irwin2010}. The authors found this to be consistent with the rapid increase of activity strength in field-age M dwarfs with spectral type around M5 \citep{West2004}, however, also note that the difference in activity indicated by the $\mathrm{H\alpha}$ line might be due to tidal interactions and subsequent spin-up of the primary by the orbiting brown dwarf.

\subsection{Chandra observations}

\noindent We observed the system on two occasions with the imaging detector of the High Resolution Camera (HRC-I) on board the {\it Chandra} X-ray Observatory. The HRC-I is sensitive to X-ray photons in the energy range from 0.08 to 10 keV but has not the capability of photon energy resolution. The FWHM for this detector is $\approx 0.4 \arcsec$, therefore, an extraction region which has a radius larger than $\approx 1 \arcsec$ will collect $>90$\% of the source photons for soft sources\footnote{See Figure 7.5 in \href{https://cxc.harvard.edu/proposer/POG/html/chap7.html}{Chapter 7} of the 'The {\it Chandra} Proposers' Observatory Guide', Version 24.0}. The two observations of NLTT~41135/41136 were taken on September 29th (Obs. ID 23388; PI Poppenhaeger) and October 1st (Obs. ID 26143; PI Poppenhaeger) 2021, both with $\approx$~25~ks exposure time.

\begin{figure}
\centering
    \includegraphics[width=\columnwidth]{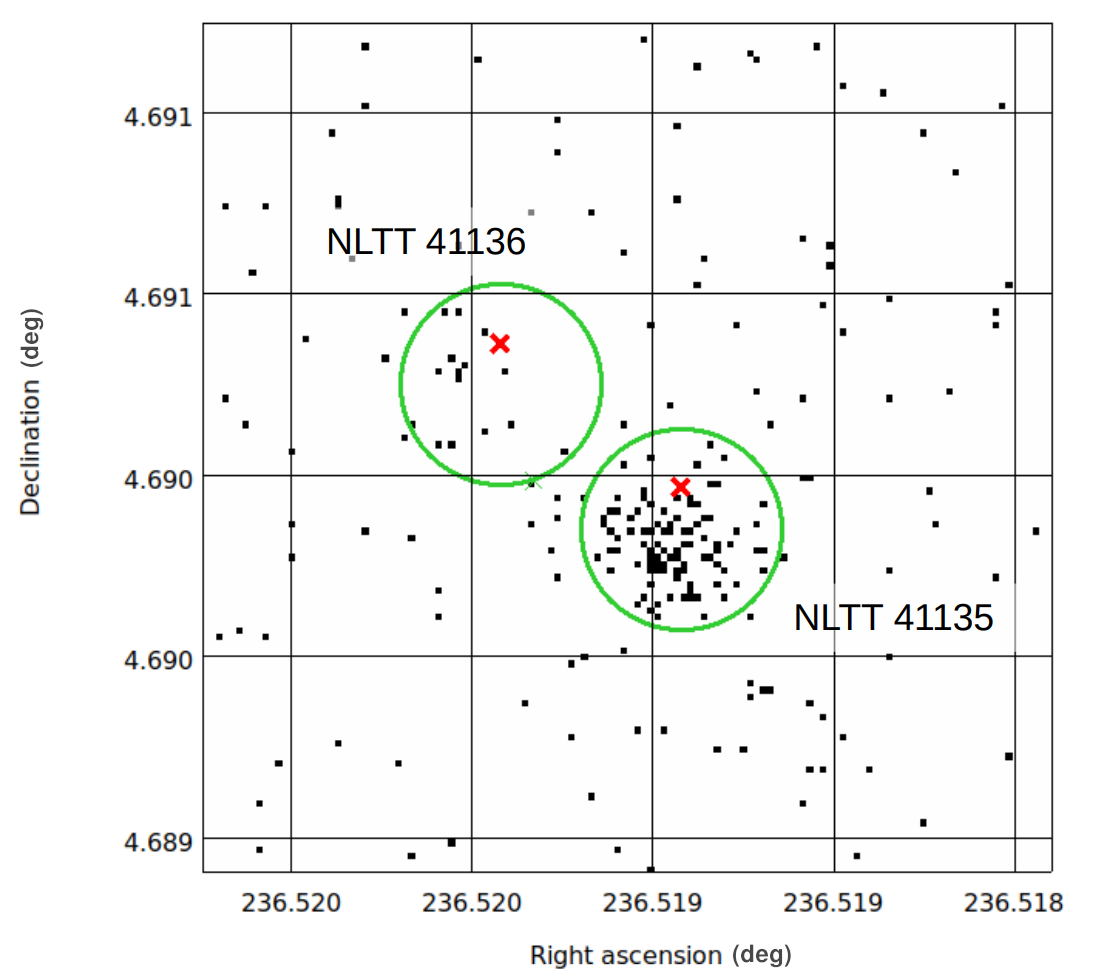}
    \caption{Combined {\it Chandra}/HRC-I observations of the system NLTT~41135/41136 with source extraction regions as green circles and the {\it Gaia} eDR3 coordinates, propagated to the epoch of observation, as red x-symbols.}
    \label{fig: NLTT41135_6}
\end{figure}

Since the stars have a projected angular separation of $2.3 \arcsec$, we chose the extraction region radius for both sources to be $1 \arcsec$, ensuring that the majority of the detected photons are collected and no overlap between the regions occurs. To estimate the background contribution to the photons in the source region, we also defined a background extraction region with a radius of $ 15 \arcsec$ in a part of the field-of-view (FoV) where the noise appears to have a uniform distribution and no astrophysical X-ray source is visible. After extracting the X-ray photons from the source regions and removing the background contribution, scaled down to the surface area of the source extraction region, we estimated the net source photon counts of the two sources. We combined the two observations to achieve a higher signal-to-noise ratio for both M dwarfs in the system (see Figure \ref{fig: NLTT41135_6}).

\begin{figure}
    \includegraphics[width=\columnwidth]{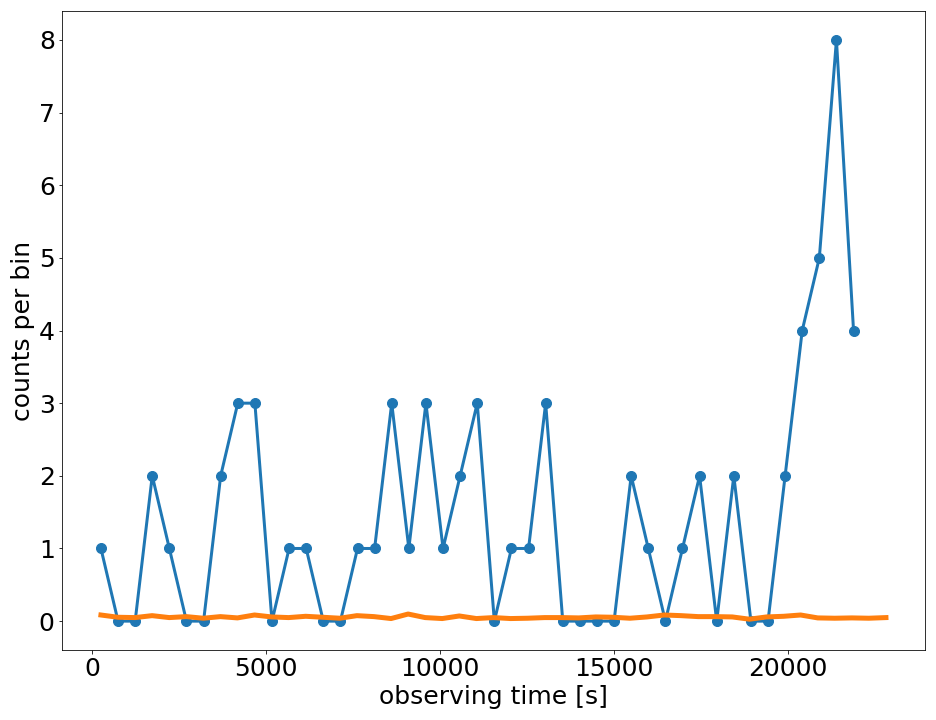}
    \caption{X-ray light curve of NLTT~41135 obtained from observation 26143. The blue dots show the count number with 500 s binsize, while the orange line shows the level of scaled background counts. At 20 ks after the observation start, a flare occurred.}
    \label{fig:LC_41135}
\end{figure}

The light curve obtained from the observation conducted in October shows a flare occurring on NLTT~41135 at $\approx$ 20 ks after the start of the observation (see Figure \ref{fig:LC_41135}). We estimated the quiescent X-ray flux by excluding the photon counts occurring in the time interval of the flare, but we also calculated the emission parameters from the whole observation time. In general, when further commenting on the activity difference between the NLTT~41135/41136 stellar components, we will be referring to the comparison of the quiescent components of the coronae. When referring to the emission with the flaring event, we will explicitly state it.

\subsection{Net source photon count}

\noindent In general, to determine the net source photon count of a faint~source, both the Poissonian uncertainties of the source and the background need to be taken into account. For the analysis of the two stars from the NLTT~41135/41136 system, we employed the analysis for faint sources described by \cite{Ilic2022}. In short, firstly, we estimated the probability that the background fluctuation was responsible for the number of counts in the source region by employing the Poisson cumulative distribution function. For both sources, we found that this probability was lower than 0.3\%, securing a $3\sigma$ level of detection significance. Furthermore, to estimate the net source photon count and its confidence interval, we applied the Kraft-Burrows-Nousek (KBN) estimator \citep{KBN1991}. The KBN estimator\footnote{In our analysis, we used the KBN implementation of the \texttt{stats.poisson\_conf\_interval} function in the \texttt{astropy} package \citep{Astropy2013,Astropy2018}.} tackles the small number statistics of faint sources in a Bayesian manner by explicitly assuming the background signal stems from a Poisson process and marginalizing over all possible background count numbers in the source detect cell. It assumes the source flux to be non-negative, and yields confidence intervals for the net source photon count. For the two detected sources, we determined the 68.3\% confidence interval and reported the number of counts at its center as the net source photon count in Table \ref{tab:nltt}.

\begin{table*}
	\centering
	\caption{Observed X-ray emission parameters for the stellar components NLTT~41135 and NLTT~41136. NLTT~41135 flared, and we estimated the parameters with (active) and without (quiescent) the flare.}
	\label{tab:nltt}
	\begin{tabular}{lccccccc} 
		\hline
        \hline
		component & counts & scaled bg counts & net counts & obs time [s] & count rate [cts/s] & $\log_{10} T$ [K] & $F_x$ $\mathrm{[erg/s/cm^2]}$\\
		\hline
		NLTT 41136 & 15 & 5.556 & $9.444_{-3.566}^{+4.240}$ & 48138 & $0.00020_{-0.00007}^{+0.00009}$ & $6.3 \pm 0.1$ & $\left(2.220_{-0.812}^{+0.964}\right)\times 10^{-15}$\\[0.1cm]
		NLTT 41135 (quiescent) & 85 & 5.244 & $79.762^{+9.565}_{-8.910}$ & 45078 & $0.00177_{-0.00020}^{+0.00021}$ & \multirow{2}{*}{$6.6 \pm 0.1$} & $\left(1.596_{-0.178}^{+0.192}\right)\times 10^{-14}$\\[0.1cm]
        NLTT 41135 (including flare) & 106 &  5.556 & $100.449_{-9.983}^{+10.643}$ & 48138 & $0.00209_{-0.00021}^{+0.00022}$ & & $\left(1.883_{-0.187}^{+0.199}\right)\times 10^{-14}$\\
		\hline
	\end{tabular}
\end{table*}

\subsection{Coronal temperature and X-ray surface flux} \label{hrctemperature}

\noindent By choosing an activity indicator from the X-ray regime, we make use of the fact that unsaturated X-ray emission is a function of stellar rotational rate: as single stars age, their rotation rate decreases, and as a consequence, their coronal temperature and X-ray emission reduce. However, if a star experiences spin-up, the average coronal temperature of the star increases, and, therefore, a higher X-ray luminosity and surface flux will be observed. 

Since the HRC-I has no intrinsic energy resolution, determining the coronal temperature directly from the observed radiation is not possible. A solution to this issue is the employment of a scaling relation between the average coronal temperature and X-ray surface flux for low-mass main-sequence stars. Observations with various X-ray telescopes have shown that the X-ray surface flux and the coronal temperature of stellar coronae are closely correlated \citep{Schmitt1997, Johnstone2015, Magaudda2022}.  

We use here the sample presented by \cite{Johnstone2015} to test where in the distribution of X-ray surface fluxes and coronal temperatures our sources would fall when assuming different coronal temperatures on a test grid. A caveat with this approach is that the coronae of M dwarfs can have multiple thermal components \citep{Schmitt1990,Giampapa1996,Robrade2005}, but having no spectral information of the observed coronae only allowed us to estimate the mean coronal temperature.

Inspired by the methodology presented by \cite{AyresBuzasi2022}, we used the online tool \href{https://cxc.harvard.edu/toolkit/pimms.jsp}{\sc webpimms} (v4.12a) to calculate expected fluxes, for given count rate, for a test grid of coronal temperatures ranging from $\log_{10} T\mathrm{[K]} = 6.0-7.0$ with 0.1 dex stepsize. The elemental abundances were set to 0.4 of the solar abundance, as suggested by \cite{Irwin2010}\footnote{\cite{Irwin2010} used the subsolar metallicity of {\it [Fe/H]} = -0.5 to model stellar parameters of the system's components. This value is between 0.2 and 0.4 solar abundances, which are the values that can be selected in {\tt webpimms}; selecting the value of 0.2 changes the resulting X-ray flux by $\approx 5\%$, which is well within the uncertainties given by the photon count confidence interval for both components.}, since NLTT~41135/41136 belongs to the old galactic disk population where the usual abundance is subsolar \citep{Leggett1992}. The temperature uncertainty is set to be equal to the stepsize of the temperature grid and influences the flux uncertainty much less than the photon count uncertainty provided in Table \ref{tab:nltt}.

\begin{figure}
    \includegraphics[width=\columnwidth]{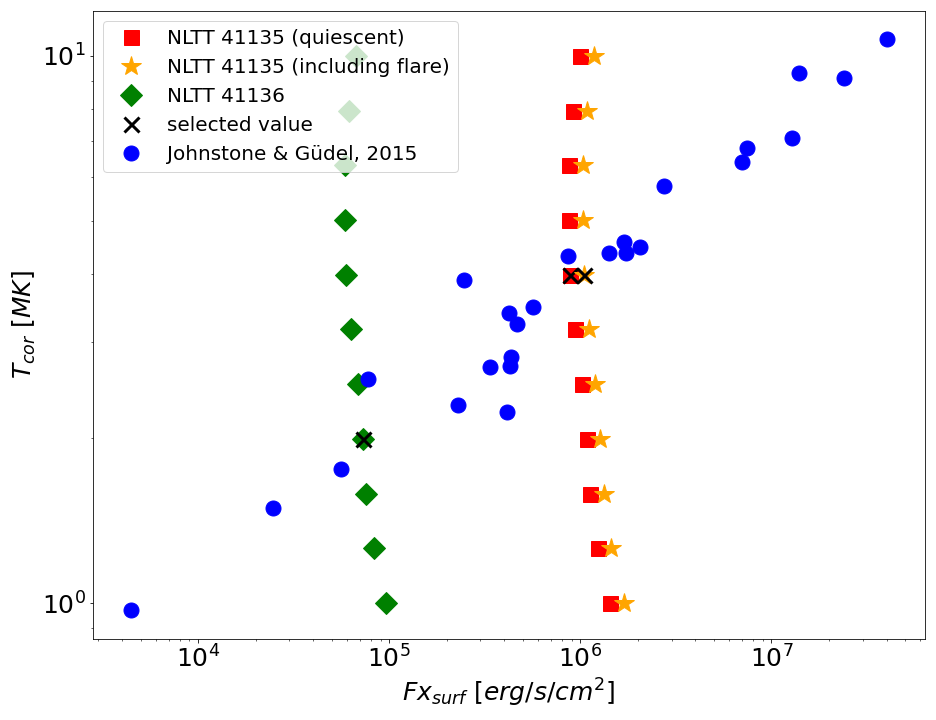}
    \caption{Coronal temperature vs. X-ray surface flux from \protect\cite{Johnstone2015} with blue dots, and the grid calculated for our targets: for NLTT~41136 with green diamond, and for NLTT~41135 with red squares and orange stars for quiescent and overall emission, respectively.}
    \label{fig:Fxsurf_Tcor}
\end{figure}

Using the resulting X-ray flux and the known distance to the system \citep{Bailer-Jones2018}, we calculated the stellar X-ray luminosity. By normalizing the luminosity with the optical surface area of the star, we arrive at the stellar X-ray surface flux value. We then compared the coronal temperature -- X-ray surface flux pairs to the sample from \cite{Johnstone2015}, which is shown in Figure~\ref{fig:Fxsurf_Tcor}. We chose the pair that matches the Johnstone sample best and proceeded with those values as our best estimates for the coronal temperature and X-ray surface flux, as given in Table~\ref{tab:nltt}.

The flux confidence interval was estimated by using the limits of the $1\sigma$ confidence interval of the net source photon count for our sources and applying these values in the \texttt{webpimms} tool with the coronal temperature estimated in the previous step. The resulting flux values were used as the upper and lower limits of the $1\sigma$ uncertainty of the X-ray flux. With error propagation, we used the estimated X-ray flux uncertainty to estimate the X-ray luminosity and -surface flux uncertainty. The distance uncertainty was not taken into account for error propagation since it is less than 1\%.

\subsection{Stellar radii}
\label{sec:radius}
\noindent Before we are able to calculate the X-ray surface flux, we needed an estimate of the stellar radius. Usually, the radii of M dwarf stars are estimated by applying the empirical relation between the Ks-band magnitude of a star and its radius, published by \cite{Mann2015}. However, the NLTT~41135/41136 system, together with several systems from the control sample, does not have a published detection in the Ks band for both components. Therefore, to have uniformly estimated stellar radii for all stars, we estimate the absolute magnitude of a star and use the main-sequence parameters published by \cite{Pecaut2013}. We employ the photometric measurement made by the ESA {\it Gaia} mission \citep{Gaia2016, Gaia2018b} and geometric distances to the stars \citep{Bailer-Jones2018} to calculate the absolute {\it G} magnitude. We then estimate the stellar radius for each star by interpolating the stellar radius vs. absolute {\it G} magnitude function for main-sequence stars\footnote{The work by \cite{Pecaut2013} is extended with the {\it Gaia} photometry for main-sequence stars in the online version: \url{https://www.pas.rochester.edu/~emamajek/EEM_dwarf_UBVIJHK_colors_Teff.txt}} for the observed magnitude.
In Table \ref{tab:photo_radii}, we give the distance, {\it Gaia} photometry, and calculated radii of all stars in our sample: the NLTT~41135/41136 system, and the control binary systems (see Section \ref{control}).

For GJ\,65, one of the control systems, the radii estimated from interferometric measurements of the stars' angular diameters \citep{Kervella2016} were available as well. We also estimate the stellar radii for this system and find a discrepancy of $\approx 10\%$ between the interferometric and our values. This is not that surprising since it was found that stellar evolutionary models, by using photometric measurements, underestimate the stellar radius of M dwarfs \citep{Ribas2006,Torres2013}. Since we are interested in the flux ratio of binary stars, we expect that, to the first order, the effect of radius discrepancy between real and model values will cancel out and will not significantly impact the final result.

\section{Results}

\subsection{The X-ray properties of NLTT~41135/41136}
\label{nltt_results}

\noindent Although the two M dwarfs in this system have similar stellar parameters, their coronal activity levels differ greatly. Firstly, from the inspection of the X-ray light curve obtained from the observation with ID 26143, a flaring event on NLTT~41135 took place towards the end. On the other hand, the primary NLTT~41136 does not show any evidence of flaring for the entire 50 ks of the observation. The quiescent count rate and the estimated average coronal temperature of the two components indicate a similar conclusion: the secondary has a count rate of $\mathrm{1.8 \times 10^{-3}}$ count/s and a coronal temperature of $\log_{10} T\mathrm{[K]}$~=~6.6, while the primary has an order of magnitude lower count rate of $\mathrm{2\times 10 ^{-4}}$ count/s and a lower temperature of $\log_{10} T \mathrm{[K]}$~=~6.3 (see Table \ref{tab:nltt} for more details). Finally, by comparing the X-ray surface fluxes of the two stars, we establish the significant difference in activity between the two stars: the brown dwarf-hosting star, NLTT~41135 with $F\mathrm{x_{surf}} = 8.9 \times 10^5$~$\mathrm{erg/s/cm^2}$, shows more than an order of magnitude greater X-ray surface flux than the primary NLTT~41136, which has a flux of $F\mathrm{x_{surf}} = 7.2 \times 10^4$~$\mathrm{erg/s/cm^2}$ (see Table \ref{tab:fluxes} for more details).

\subsection{M dwarf wide binary systems for activity difference comparison}
\label{control}

\noindent To determine the significance of the measured coronal activity difference between NLTT~41135 and NLLT~41136, and, therefore, the tidal influence of the brown dwarf, we constructed a comparison sample of wide M dwarf binaries without known close-in substellar companions. We searched the \cite{El-Badry2018} wide stellar binaries catalog for M dwarf pairs that harbor stars with similar {\it Gaia} $G-G_{\mathrm{RP}}$ color. We selected pairs where the absolute value of the color difference is $|d(G-G_{\mathrm{RP}})| \leq 0.2$, i.e. where the stellar components can differ by up to three spectral subtypes. We expect that coeval M dwarfs with colors differing by more than the equivalent of three spectral subtypes will follow somewhat different stellar evolutionary paths, leading the stars to display noticeably different activity levels in the X-ray regime. This scenario might be true for some coeval stars differing by $|d(G-G_{\mathrm{RP}})| \leq 0.2$; however, the selected color-difference limit is a compromise between having a significant number of systems for comparison and them having intrinsically different stars. Aside from the color-difference constraint, we also imposed a constraint on the distance to the systems since most moderately active M dwarfs will not be detected in routine X-ray observations at larger distances. Below, we describe the query for M dwarf pairs in the source catalogs and data archives of three X-ray space observatories: the {\it XMM-Newton} space telescope, the {\it Chandra} X-ray Observatory, and the eROSITA instrument onboard the Spectrum R\"ontgen Gamma satellite.

\subsubsection{XMM-Newton}

To find M dwarf pairs observed with {\it XMM-Newton}'s European Photon Imaging Camera (EPIC), we used input pairs from the \cite{El-Badry2018} sample as described above and considered pairs within a distance of 100\,pc and where the two stars have a separation of at least $15 \arcsec$, since stars at closer separation are typically spatially blended in EPIC observations. We used the 4XMM-DR10 catalog, version 1.0 \citep{Webb2020} for cross-matching. Since \textit{XMM-Newton} observations have been conducted since 1999 and several of our nearby input pairs display significant proper motions, we first searched for potential matches between the \cite{El-Badry2018} sample with coordinates evolved to the J2010.0 epoch and the {\it XMM-Newton} catalog with a large matching radius of 30\arcsec. We then extracted the actual observational epoch(s) from the {\it XMM-Newton} catalog, evolved the stellar coordinates to that specific epoch, or in case of multiple {\it XMM-Newton} observations to the average epoch, and refined the cross-match by using a matching radius of $5\arcsec$ with the updated coordinates. A visual inspection of the resulting matches in ESASky\footnote{\url{https://sky.esa.int/esasky/}, \cite{Giordano2018}} showed that for some pairs there was source confusion present, especially when the targets were located at the edge of the EPIC field of view. We, therefore, proceeded by downloading the individual observational data sets from the {\it XMM-Newton} archive and performed a customized analysis by hand.

The query of the 4XMM-DR10 catalog resulted in 10 M dwarf systems. One of these systems, BX Tri, is a hierarchical system hosting a tight pair that was not flagged as such in the \cite{El-Badry2018} catalog, and which is not spatially resolved by {\it XMM-Newton}. We analysed the observations of the remaining nine systems, obtained from the \href{https://nxsa.esac.esa.int/nxsa-web/\#search}{XMM-Newton data archive}, as described by \cite{Ilic2022}, and find six systems where both components are detected, and three systems where one component is not detected. We report the fluxes of the detected M dwarfs, and the upper limits\footnote{We estimated the upper limit of the flux values for all undetected M dwarfs in the control sample by calculating the 99.7\% ($3\sigma$) one-sided confidence interval of the underlying Poissonian distribution of photons found in a suitable background-region, scaled down to the size of the source extraction region.} of the undetected ones in Table \ref{tab:fluxes}. There, the flux values are given for the energy range of 0.2-2.0 keV, where stellar coronae typically emit the bulk of their radiation \citep{Gudel1997}. The values of the X-ray fluxes are computed assuming an underlying thermal spectrum with a coronal temperature estimated from the observed hardness ratio, as described in more detail by \cite[see Appendix \ref{AppB} for the details on individual systems]{Ilic2022}.

\subsubsection{Chandra}

The {\it Chandra} X-ray observatory has a higher spatial resolution, but generally lower sensitivity than {\it XMM-Newton}. We, therefore, selected M dwarf pairs with a spatial separation of at least 2\arcsec and within a distance of 50\,pc from the \cite{El-Badry2018} sample and cross-matched them with the  {\it Chandra} Source Catalog 2.0 (CSC2.0, \citealt{Evans2018}) in a similar manner as we did for the {\it XMM-Newton}. Again, since  {\it Chandra} has been in operation since 1999, we used stellar coordinates evolved to an epoch of J2010.0 for initial cross-matching with a matching radius of 30\arcsec. We then directly proceeded by downloading the individual observation files from the  {\it Chandra} archive\footnote{\url{https://cda.harvard.edu/chaser/}} and performed source detection analysis as described by \cite{Ilic2022}. We found two systems where both components are detected, three systems where no component was detected, and three systems where one component is not detected; for all undetected stars, we again report the one-sided $3\sigma$ confidence interval as the upper limit on the X-ray flux. Out of those eight systems, all but one were observed with the ACIS instrument, and we estimated energy conversion factors as described by \cite{Ilic2022} from hardness ratios and reported X-ray fluxes in the 0.2-2.0~keV energy band. For the one system that was observed with the HRC instrument, we perform the same analysis as described in section~\ref{hrctemperature} for the NLTT41135/41136 system to find its X-ray flux in the full HRC energy band of 0.08-10~keV. In Appendix \ref{AppB} we provide more details on individual systems together with the observed radiation hardness ratio and estimated coronal temperature used for the X-ray flux calculation.

We also added the known binary system GJ~65 to the control sample, which was not included in the  wide stellar binary catalog by \cite{El-Badry2018}. A detailed analysis of the two recent observations (obs ID: 22344 and 22876) is performed by Wolk et al.\ (in prep) and the resulting fluxes are included here. The X-ray photometry of M dwarfs observed with {\it Chandra} is reported in Table \ref{tab:fluxes}.

\subsubsection{eROSITA M dwarf pairs}

\noindent We compared the selected M dwarf wide binaries from the catalog described above to the catalog of X-ray sources detected with the R\"ontgen Survey with an Imaging Telescope Array (eROSITA) \citep{Merloni2012,Freyberg2020,Dennerl2020,Predehl2021,Brunner2022}, an X-ray instrument onboard the Russian Spectrum-R\"ontgen-Gamma spacecraft \citep{Sunyaev2021,Pavlinsky2021}. It was launched in mid-2019 into an orbit around the $\mathrm{L_2}$ Lagrange point of the Sun-Earth system. eROSITA consists of seven Wolter telescopes with one camera assembly each and is sensitive to photon energies between 0.2-10 keV \citep{Meidinger2020}. eROSITA started an all-sky survey in 2019, where it scans the whole sky every six months in great circles roughly perpendicular to the ecliptic. Any point on the sky is scanned every four hours for several eROSITA slews, with the number of slews when a given target is in the field of view depending on the ecliptic latitude of the target. 

eROSITA has completed four all-sky surveys to date (named eRASS1 to eRASS4), as well as a partially-completed fifth all-sky survey (eRASS5). In addition to source catalogs from each of those surveys, the eROSITA\_DE consortium has also produced a catalog from the stacked data of the four completed eRASS surveys in the German part of the eROSITA sky, called eRASS:4, accessible within the eROSITA consortium in the data reduction version from October 31 2022. eROSITA has wide wings of its PSF; to avoid issues with source blending and upper limit calculations, we selected known M dwarf binary pairs within a volume of 50 pc that have a separation of at least $50\arcsec$ between the two stars. We then matched those individual stars to the catalog with a matching radius of $10 \arcsec$ and checked that all matched X-ray sources are likely to be of stellar nature, as described by \citet{Foster2022}.

The stacked eRASS:4 survey (as well as the individual eRASS surveys) is shallow, with total exposure times of the order of 500\,s. M dwarfs frequently produce X-ray flares, which increases the probability that some of the M dwarf detections in eRASS:4 were only achieved because the M dwarf flared during the exposure time. Indeed, there is evidence of this reported in \cite{Stelzer2022}. Since we want to compare the quiescent X-ray emission levels of stars, we therefore clean the initially matched sample as follows: we require that a given star, in addition to being detected in the stacked eRASS:4 survey, is detected in at least three of the five individual eRASS1 to eRASS5 surveys, meaning we have likely seen the quiescent emission from the star. We then estimate the quiescent flux of the star by taking the median of the individually detected flux values, and its uncertainty by the standard deviation of those detected individual fluxes. We note that three detections is the minimum number required to identify one outlier and determine the typical flux via the median.

Systems in which fewer than three flux detections were achieved were considered to be dominated by flaring emission and, since we do not have a way to characterize the quiescent flux from the available data, they were discarded. Systems in which one star had a detected quiescent flux as described above, but the other star had no detection, were kept in the sample and an upper limit to the flux of the undetected star was computed using the prescription of Tub\'in-Arenas et al.\ 2023 (submitted), i.e.\ by performing X-ray photometry on the eROSITA standard calibration data products (counts image, background image, and exposure time), following the Bayesian approach described by
\citet{Kraft1991}. The upper limits are given as one-sided $3\sigma$ confidence intervals in the eROSITA soft band, which has an energy range of 0.2-2.3 keV, and using appropriate energy conversion factors for a stellar corona with a temperature of $kT=0.3$\,keV, typical for moderately active stars \citep{Schmitt1990,Foster2022}.
This procedure yielded ten pairs where both stars have a detected quiescent flux and five pairs where one star has a detected quiescent flux and the other star has an upper limit.

\begin{figure}
    \includegraphics[width=\columnwidth]{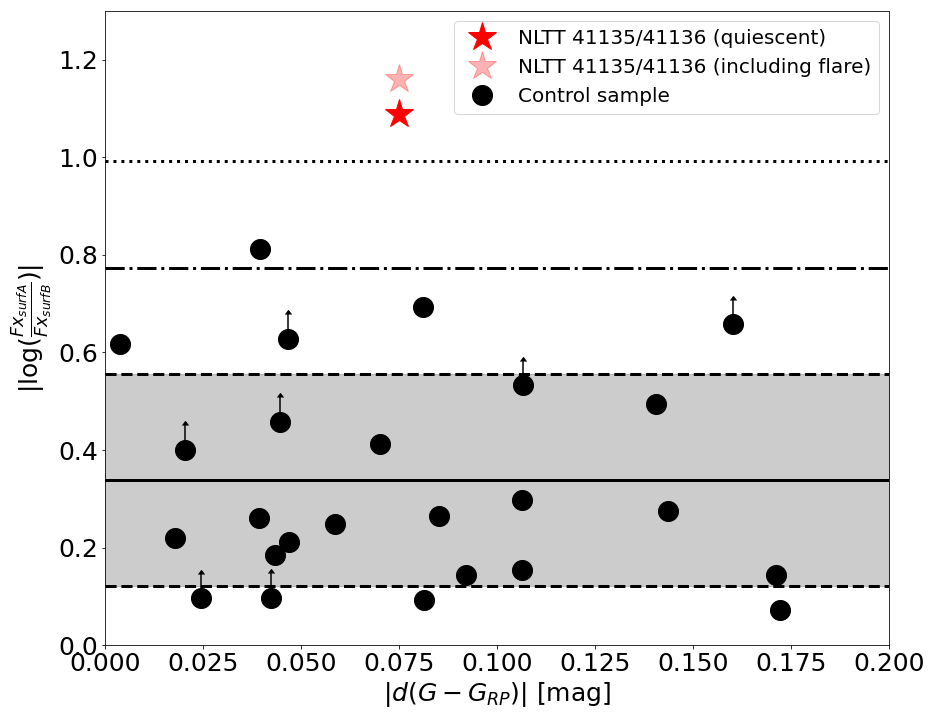}
    \caption{The activity difference as a function of the absolute value of color difference, which corresponds to the mass difference between the components of the wide binary. The binary systems from the control sample are shown as black dots and the NLTT~41135/41136 system as the red asterisk symbol, with the fainter red asterisk including the flaring episode. The solid black line is the mean of the activity difference of the control sample, the shaded area indicates the $1\sigma$ confidence interval around the mean, while the dashed and dotted lines represent the $+2\sigma$ and $+3\sigma$ confidence interval limits, respectively.}
    \label{fig:gaia_comp}
\end{figure}

\begin{figure}
    \includegraphics[width=\columnwidth]{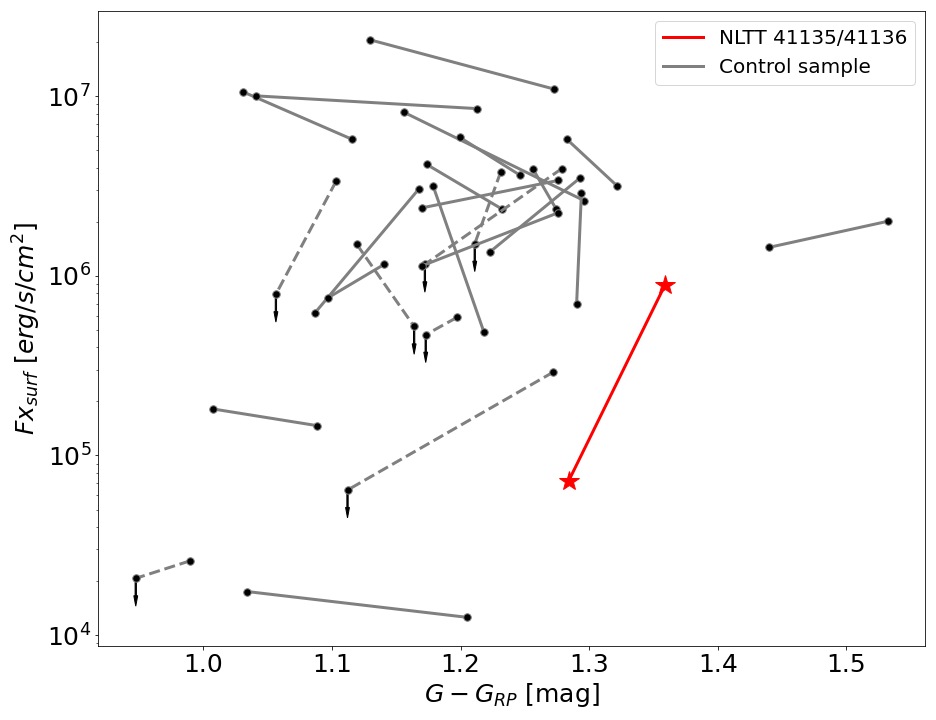}
    \caption{The X-ray surface flux as a function of the {\it Gaia} color of M dwarfs in the control sample and the NLTT~41135/41136 system. The two components of a binary are connected with a line: the gray solid line shows binaries where both components are detected, the gray dashed line shows systems where one component is undetected, and the red line connects NLTT~41135 and NLTT~41136.}
    \label{fig:control_sample}
\end{figure}

\onecolumn
\begin{landscape}
\begin{table*}
\caption{Spectral type (SpT) and projected physical separation ($\rho$) of the components together with their {\it Gaia} DR2 photometry and distances \protect\citep{Bailer-Jones2018} used to calculate the absolute G magnitudes and the radii. Additionally, the absolute value of the color difference between the binary components and their {\it Gaia} DR2 stellar proper motion, used to evolve coordinates to a certain epoch, are given.}
	\label{tab:photo_radii}
\begin{tabular}{llcccccccccc}

\hline
\hline
                     component & SpT &             $\rho$ [AU] &  dist [pc] &       G &  $G-G_{RP}$ &   $M_G$ &  $R_{Gaia}[R_{\odot}]$ &         $|d(G-G_{RP})|$  & $\mu_{\alpha} \cdot\cos\delta$[mas/yr] & $\mu_{\delta}$[mas/yr] \\
\hline
        NLTT 41136 & M4 & \multirow{2}{*}{79} & 34.319 & 13.966 & 1.284 & 11.289 & 0.269 & \multirow{2}{*}{0.075} & $153.670 \pm 0.243$ & $-281.981 \pm 0.241$\\
        NLTT 41135 & M5 & & 34.136 & 14.936 & 1.359 & 12.270 & 0.205 & & $162.509 \pm 0.182$ & $-282.725 \pm 0.182$\\[0.1cm]
                      GJ 15 B &  M3 &    \multirow{2}{*}{122} &      3.561 &   9.677 &       1.205 &  11.919 &                  0.225 &  \multirow{2}{*}{0.171} &                    $2863.284\pm0.069$ &      $336.529\pm0.039$ \\
                      GJ 15 A &  M1 &                      &      3.562 &   7.216 &       1.034 &   9.458 &                  0.435 &                      &                    $2891.525\pm0.061$ &      $411.903\pm0.034$ \\[0.1cm]
 Gaia DR2 1608710752684301312 &  M4 &   \multirow{2}{*}{4908} &     74.594 &  15.646 &       1.294 &  11.283 &                  0.269 &  \multirow{2}{*}{0.004} &                     $-27.898\pm0.095$ &       $30.127\pm0.084$ \\
 Gaia DR2 1608710791338814208 &  M4 &                      &     75.330 &  16.155 &       1.290 &  11.770 &                  0.236 &                      &                     $-29.958\pm0.103$ &       $29.634\pm0.085$ \\[0.1cm]
                     Ross 868 &  M3 &    \multirow{2}{*}{181} &     10.751 &  10.137 &       1.170 &   9.980 &                  0.372 &  \multirow{2}{*}{0.106} &                    $-214.777\pm0.077$ &      $351.001\pm0.084$ \\
                     Ross 867 &  M4 &                      &     10.753 &  11.456 &       1.276 &  11.298 &                  0.268 &                      &                    $-226.132\pm0.098$ &      $355.284\pm0.105$ \\[0.1cm]
                TIC 293303829 &  M1 &   \multirow{2}{*}{4576} &     55.871 &  12.230 &       1.019 &   8.494 &                  0.538 &  \multirow{2}{*}{0.023} &                     $-24.591\pm0.117$ &     $-167.143\pm0.057$ \\
                TIC 293303832 &  M0 &                      &     55.853 &  12.063 &       0.996 &   8.328 &                  0.562 &                      &                     $-24.573\pm0.095$ &     $-166.795\pm0.046$ \\[0.1cm]
 Gaia DR2 3074577322667614976 &  M3 &  \multirow{2}{*}{17201} &     53.705 &  13.730 &       1.199 &  10.080 &                  0.359 &  \multirow{2}{*}{0.047} &                     $-43.033\pm0.092$ &      $-86.589\pm0.065$ \\
 Gaia DR2 3074630580262065536 &  M4 &                      &     53.625 &  14.777 &       1.246 &  11.130 &                  0.280 &                      &                     $-42.529\pm0.106$ &      $-86.487\pm0.066$ \\[0.1cm]
                TIC 436632332 &  M4 &   \multirow{2}{*}{2176} &     83.799 &  14.970 &       1.282 &  10.353 &                  0.338 &  \multirow{2}{*}{0.039} &                      $19.706\pm0.148$ &      $-43.233\pm0.086$ \\
                TIC 436632331 &  M4 &                      &     83.769 &  14.970 &       1.322 &  10.355 &                  0.338 &                      &                      $20.083\pm0.141$ &      $-43.467\pm0.083$ \\[0.1cm]
 Gaia DR2 4768120070358120064 &  M4 &   \multirow{2}{*}{3219} &     91.156 &  16.887 &       1.272 &  12.088 &                  0.215 &   \multirow{2}{*}{0.16} &                      $37.732\pm0.156$ &       $11.151\pm0.182$ \\
 Gaia DR2 4768120104717857792 &  M2 &                      &     91.011 &  14.854 &       1.112 &  10.058 &                  0.360 &                      &                      $37.515\pm0.078$ &       $10.981\pm0.099$ \\[0.1cm]
                     G 202-66 &  M3 &   \multirow{2}{*}{1526} &     60.676 &  13.395 &       1.179 &   9.480 &                  0.433 &  \multirow{2}{*}{0.039} &                     $-239.59\pm0.058$ &      $174.561\pm0.064$ \\
                     G 202-67 &  M3 &                      &     60.676 &  15.046 &       1.218 &  11.131 &                  0.280 &                      &                     $-241.61\pm0.072$ &      $175.549\pm0.089$ \\[0.1cm]
                     LTT 6326 &  M0 &   \multirow{2}{*}{1640} &     61.690 &  12.796 &       0.990 &   8.845 &                  0.498 &  \multirow{2}{*}{0.042} &                     $-55.483\pm0.037$ &     $-334.752\pm0.042$ \\
                     LTT 6325 &  M0 &                      &     61.774 &  12.397 &       0.948 &   8.443 &                  0.544 &                      &                     $-55.444\pm0.038$ &     $-333.655\pm0.042$ \\[0.1cm]
                                         LP320-163 &  M6 &   \multirow{2}{*}{102} &     45.653 &  16.098 &       1.447 &  12.801 &                  0.180 &  \multirow{2}{*}{0.032} &                    $-212.555\pm0.155$ &        $27.959\pm0.12$ \\
                    LP320-162 &  M5 &                     &     46.370 &  16.152 &       1.416 &  12.821 &                  0.180 &                      &                    $-218.427\pm0.153$ &       $28.591\pm0.118$ \\[0.1cm]
                   LP920-61 B &  M3 &    \multirow{2}{*}{87} &     18.234 &  12.506 &       1.197 &  11.202 &                  0.275 &  \multirow{2}{*}{0.024} &                    $-121.425\pm0.105$ &     $-122.807\pm0.075$ \\
                   LP920-61 A &  M3 &                     &     18.268 &  12.450 &       1.173 &  11.141 &                  0.279 &                      &                    $-113.453\pm0.106$ &     $-123.013\pm0.075$ \\[0.1cm]
             SCR J0602-3952-B &  M3 &   \multirow{2}{*}{126} &     45.719 &  14.274 &       1.232 &  10.973 &                  0.292 &  \multirow{2}{*}{0.054} &                      $75.063\pm0.073$ &       $178.107\pm0.07$ \\
             SCR J0602-3952-A &  M3 &                     &     45.885 &  13.436 &       1.178 &  10.127 &                  0.355 &                      &                      $72.687\pm0.051$ &      $175.841\pm0.049$ \\[0.1cm]
                 TIC 20446899 &  M0 &  \multirow{2}{*}{1778} &     40.659 &  11.402 &       0.969 &   8.356 &                  0.557 &  \multirow{2}{*}{0.184} &                    $-170.787\pm0.093$ &         $4.69\pm0.068$ \\
 Gaia DR2 3532611086293698560 &  M3 &                     &     40.566 &  13.127 &       1.152 &  10.086 &                  0.358 &                      &                    $-172.895\pm0.096$ &        $7.383\pm0.066$ \\[0.1cm]
                       G236-1 &  M2 &  \multirow{2}{*}{1054} &     11.850 &   9.563 &       1.097 &   9.195 &                  0.457 &  \multirow{2}{*}{0.043} &                    $-671.954\pm0.045$ &     $-271.231\pm0.058$ \\
                       G236-2 &  M3 &                     &     11.873 &  10.067 &       1.141 &   9.695 &                  0.417 &                      &                    $-671.125\pm0.055$ &     $-265.525\pm0.062$ \\[0.1cm]
  Gaia DR2 715928515183511040 &  M4 &   \multirow{2}{*}{387} &     36.055 &  14.404 &       1.275 &  11.619 &                  0.246 &  \multirow{2}{*}{0.118} &                       $4.986\pm0.101$ &       $53.474\pm0.096$ \\
                 TIC 16151129 &  M3 &                     &     35.905 &  12.839 &       1.157 &  10.063 &                  0.360 &                      &                        $3.271\pm0.08$ &         $58.59\pm0.07$ \\[0.1cm]
                   Ross 110 B &  M4 &    \multirow{2}{*}{68} &     21.557 &  13.196 &       1.262 &  11.529 &                  0.252 &  \multirow{2}{*}{0.014} &                     $544.582\pm0.062$ &     $-549.415\pm0.062$ \\
                   Ross 110 A &  M4 &                     &     21.479 &  12.904 &       1.248 &  11.244 &                  0.272 &                      &                      $547.56\pm0.125$ &     $-535.855\pm0.119$ \\[0.1cm]
                   GJ 65 A & M6 & \multirow{2}{*}{6} & 2.703 &  10.507 & 1.440 & 13.348 & 0.156 & \multirow{2}{*}{0.092} & $3385.827 \pm 0.492$ & $532.040 \pm 0.374$ \\
                   GJ 65 B & M7 & &  2.687 &  10.869 & 1.532 & 13.723 & 0.148 & & $3182.734 \pm 0.552$ & $592.104 \pm 0.427$\\
             
\hline
\end{tabular}
\end{table*}
\end{landscape}
\twocolumn

\onecolumn
\begin{landscape}
\addtocounter{table}{-1}
\begin{table*}
\caption{Continued}
\begin{tabular}{llcccccccccc}

\hline
\hline
                     component & SpT &             $\rho$ [AU] &  dist [pc] &       G &  $G-G_{RP}$ &   $M_G$ &  $R_{Gaia}[R_{\odot}]$ &         $|d(G-G_{RP})|$  & $\mu_{\alpha} \cdot\cos\delta$[mas/yr] & $\mu_{\delta}$[mas/yr] \\
\hline
              TIC 206617096 &  M4 &  \multirow{2}{*}{18438} &     34.690 &  14.375 &       1.274 &  11.674 &                  0.242 &  \multirow{2}{*}{0.018} &                       $1.598\pm0.077$ &       $84.663\pm0.075$ \\
                TIC 206617113 &  M4 &                      &     34.703 &  14.168 &       1.257 &  11.467 &                  0.256 &                      &                       $1.177\pm0.063$ &       $85.017\pm0.056$ \\[0.1cm]
                      GJ 3148 &  M3 &   \multirow{2}{*}{1415} &     13.408 &  10.777 &       1.174 &  10.140 &                  0.354 &  \multirow{2}{*}{0.059} &                     $684.129\pm0.127$ &      $248.772\pm0.103$ \\
                      GJ 3149 &  M3 &                      &     13.433 &  11.749 &       1.232 &  11.108 &                  0.282 &                      &                     $683.716\pm0.161$ &      $246.394\pm0.134$ \\[0.1cm]
             UCAC4 235-004550 &  M1 &   \multirow{2}{*}{2601} &     39.520 &  11.739 &       1.031 &   8.755 &                  0.508 &  \multirow{2}{*}{0.085} &                        $35.9\pm0.043$ &      $-33.357\pm0.061$ \\
             UCAC4 235-004546 &  M2 &                      &     39.551 &  12.384 &       1.116 &   9.398 &                  0.439 &                      &                      $37.114\pm0.048$ &      $-32.231\pm0.078$ \\[0.1cm]
 Gaia DR2 4899032116649119616 &  M4 &  \multirow{2}{*}{13801} &     38.939 &  14.766 &       1.269 &  11.814 &                  0.233 &  \multirow{2}{*}{0.121} &                     $-28.019\pm0.073$ &      $-45.644\pm0.077$ \\
 Gaia DR2 4899029951985608576 &  M5 &                      &     39.098 &  16.142 &       1.391 &  13.181 &                  0.164 &                      &                     $-28.182\pm0.109$ &      $-46.464\pm0.113$ \\[0.1cm]
                TIC 167422188 &  M1 &  \multirow{2}{*}{13368} &     40.774 &  11.890 &       1.007 &   8.839 &                  0.499 &  \multirow{2}{*}{0.081} &                      $79.093\pm0.048$ &      $234.317\pm0.046$ \\
                TIC 167417695 &  M2 &                      &     41.058 &  12.628 &       1.089 &   9.561 &                  0.428 &                      &                      $78.676\pm0.053$ &      $233.277\pm0.054$ \\[0.1cm]
                TIC 106344480 &  M1 &  \multirow{2}{*}{45353} &     24.390 &  10.606 &       1.041 &   8.670 &                  0.518 &  \multirow{2}{*}{0.172} &                      $37.481\pm0.038$ &      $-34.252\pm0.039$ \\
                TIC 106493402 &  M3 &                      &     24.466 &  12.216 &       1.213 &  10.274 &                  0.344 &                      &                       $36.45\pm0.051$ &       $-31.75\pm0.067$ \\[0.1cm]
                TIC 416857959 &  M2 &   \multirow{2}{*}{5204} &     39.125 &  12.356 &       1.103 &   9.393 &                  0.439 &  \multirow{2}{*}{0.047} &                     $-36.339\pm0.051$ &       $47.232\pm0.054$ \\
                TIC 450297524 &  M1 &                      &     39.122 &  12.097 &       1.056 &   9.135 &                  0.464 &                      &                     $-36.251\pm0.051$ &        $48.66\pm0.048$ \\[0.1cm]
                 TIC 36765037 &  M3 &   \multirow{2}{*}{2631} &     30.149 &  13.589 &       1.231 &  11.192 &                  0.275 &   \multirow{2}{*}{0.02} &                      $-183.1\pm0.462$ &      $-26.428\pm0.697$ \\
                 TIC 36765044 &  M3 &                      &     31.660 &  13.626 &       1.211 &  11.124 &                  0.281 &                      &                     $-182.39\pm0.152$ &      $-25.195\pm0.129$ \\[0.1cm]
                TIC 151639642 &  M2 &   \multirow{2}{*}{3799} &     36.792 &  12.031 &       1.087 &   9.202 &                  0.456 &  \multirow{2}{*}{0.081} &                      $31.746\pm0.043$ &      $-79.325\pm0.045$ \\
             UCAC4 270-056947 &  M3 &                      &     36.838 &  12.738 &       1.168 &   9.906 &                  0.384 &                      &                      $31.723\pm0.049$ &       $-78.358\pm0.05$ \\[0.1cm]
             UCAC4 150-081944 &  M4 &   \multirow{2}{*}{6132} &     30.374 &  14.268 &       1.296 &  11.855 &                  0.230 &   \multirow{2}{*}{0.14} &                     $-145.97\pm0.097$ &       $-64.886\pm0.08$ \\
                TIC 317385747 &  M3 &                      &     30.317 &  12.357 &       1.156 &   9.949 &                  0.377 &                      &                    $-147.448\pm0.054$ &      $-66.219\pm0.047$ \\[0.1cm]
                TIC 392786054 &  M3 &   \multirow{2}{*}{9659} &     26.342 &  12.641 &       1.223 &  10.538 &                  0.325 &   \multirow{2}{*}{0.07} &                     $159.768\pm0.113$ &      $-83.222\pm0.098$ \\
                TIC 392785968 &  M4 &                      &     26.445 &  13.618 &       1.293 &  11.506 &                  0.254 &                      &                     $160.361\pm0.124$ &      $-82.213\pm0.098$ \\[0.1cm]
                      WT 2090 &  M4 &   \multirow{2}{*}{5413} &     21.426 &  13.371 &       1.279 &  11.716 &                  0.239 &  \multirow{2}{*}{0.107} &                    $-171.652\pm0.125$ &      $-247.082\pm0.11$ \\
                    Wolf 1501 &  M3 &                      &     21.408 &  12.113 &       1.172 &  10.461 &                  0.330 &                      &                    $-170.941\pm0.093$ &     $-247.728\pm0.081$ \\[0.1cm]
                TIC 410458113 &  M2 &   \multirow{2}{*}{4370} &     27.554 &  11.787 &       1.119 &   9.586 &                  0.427 &  \multirow{2}{*}{0.045} &                      $88.314\pm0.119$ &      $-98.339\pm0.082$ \\
             UCAC4 385-070621 &  M3 &                      &     27.579 &  12.656 &       1.164 &  10.453 &                  0.331 &                      &                      $89.838\pm0.121$ &      $-95.569\pm0.082$ \\[0.1cm]
                TIC 229807000 &  M2 &  \multirow{2}{*}{35099} &     45.987 &  11.983 &       1.129 &   8.670 &                  0.518 &  \multirow{2}{*}{0.144} &                      $73.503\pm0.041$ &      $-68.137\pm0.042$ \\
                TIC 229807051 &  M4 &                      &     45.885 &  13.875 &       1.273 &  10.566 &                  0.323 &                      &                      $73.624\pm0.071$ &      $-68.011\pm0.071$ \\
\hline
\end{tabular}
\end{table*}
\end{landscape}

\onecolumn
\begin{landscape}
\begin{table*}
	\caption{X-ray parameters for all the stars in our sample, together with the activity difference in each binary. Additionally, we provide the instrument with which {\it Chandar} systems were observed; {\it XMM-Newton} systems were observed only with the EPIC instrument.}
	\label{tab:fluxes}

\begin{tabular}{llccccccc}
\hline
\hline
                     mission &                     component & $F_x \times 10^{-14}$ $[erg/s/cm^2]$ & $L_x \times 10^{27}$ [erg/s] & $Fx_{surf} \times 10^5$ $[erg/s/cm^2]$ & $L_{bol} \times 10^{31}$ [erg/s] &  $\log R_x$ & $\left|\log \frac{Fx_{surfA}}{Fx_{surfB}}\right|$ \\
\hline

\multirow{3}{*}{Chandra/HRC} & NLTT 41136 & $0.22 \pm 0.09$ & $0.32 \pm 0.13$ & $0.72 \pm 0.29$ & 2.61 & -4.91 & \multirow{2}{*}{$1.09 \pm 0.18$}\\
                             & NLTT 41135 (quiescent) & $1.6 \pm 0.2$ & $2.3 \pm 0.3$ & $ 8.9 \pm 1.0$ & \multirow{2}{*}{$1.31 \times 10^{31}$} & -3.76 &\\
                             & NLTT 41135 (average) & $1.9 \pm 0.2$ & $ 2.7 \pm 0.3 $ & $10.0 \pm 1.0$ & & -3.69 & $1.16 \pm 0.18$\\[0.1cm]

\multirow{2}{*}{XMM-Newton} &                       GJ 15 B &                  $2.506 \pm 0.136$ &              $0.04 \pm 0.0$ &                        $0.13 \pm 0.01$ &                           $1.67$ & -5.63 &      \multirow{2}{*}{$0.14 \pm 0.03$} \\
                             &                       GJ 15 A &                 $12.992 \pm 0.285$ &               $0.2 \pm 0.0$ &                         $0.17 \pm 0.0$ &                          $10.02$ & -5.70 &                                       \\[0.1cm]
 \multirow{2}{*}{XMM-Newton} &  Gaia DR2 1608710752684301312 &                   $1.88 \pm 0.144$ &            $12.76 \pm 0.98$ &                        $29.0 \pm 2.23$ &                           $2.64$ & -3.32 &      \multirow{2}{*}{$0.62 \pm 0.09$} \\
                             &  Gaia DR2 1608710791338814208 &                  $0.341 \pm 0.068$ &             $2.36 \pm 0.47$ &                         $7.0 \pm 1.39$ &                           $1.86$ & -3.90 &                                       \\[0.1cm]
 \multirow{2}{*}{XMM-Newton} &                      Ross 868 &                 $67.726 \pm 0.802$ &             $9.55 \pm 0.11$ &                       $11.34 \pm 0.13$ &                           $6.65$ & -3.84 &       \multirow{2}{*}{$0.3 \pm 0.01$} \\
                             &                      Ross 867 &                 $69.566 \pm 0.815$ &             $9.81 \pm 0.11$ &                       $22.47 \pm 0.26$ &                           $2.62$ & -3.43 &                                       \\[0.1cm]
 \multirow{2}{*}{XMM-Newton} &                 TIC 293303829 &                  $0.178 \pm 0.055$ &             $0.68 \pm 0.21$ &                        $0.38 \pm 0.12$ &                          $19.86$ & -5.47 &   \multirow{2}{*}{/}    \\
                             &                 TIC 293303832 &                    $\leq 0.451     $ &              $\leq 1.72     $ &                         $\leq 0.89     $ &                          $22.85$ & $\leq -5.12$ &                                       \\[0.1cm]
 \multirow{2}{*}{XMM-Newton} &  Gaia DR2 3074577322667614976 &                 $13.115 \pm 0.219$ &            $46.14 \pm 0.77$ &                       $58.93 \pm 0.98$ &                           $6.11$ & -3.12 &      \multirow{2}{*}{$0.21 \pm 0.02$} \\
                             &  Gaia DR2 3074630580262065536 &                  $4.934 \pm 0.152$ &            $17.31 \pm 0.53$ &                       $36.26 \pm 1.12$ &                           $2.96$ & -3.23 &                                       \\[0.1cm]
 \multirow{2}{*}{XMM-Newton} &                 TIC 436632332 &                  $4.698 \pm 0.272$ &            $40.25 \pm 2.33$ &                       $57.77 \pm 3.34$ &                           $5.08$ & -3.10 &      \multirow{2}{*}{$0.26 \pm 0.04$} \\
                             &                 TIC 436632331 &                  $2.582 \pm 0.197$ &             $22.1 \pm 1.69$ &                       $31.75 \pm 2.42$ &                           $5.08$ & -3.36 &                                       \\[0.1cm]
 \multirow{2}{*}{XMM-Newton} &  Gaia DR2 4768120070358120064 &                   $0.08 \pm 0.021$ &             $0.82 \pm 0.21$ &                        $2.91 \pm 0.76$ &                           $1.48$ & -4.26 &       \multirow{2}{*}{$\geq 0.66     $} \\
                             &  Gaia DR2 4768120104717857792 &                     $\leq0.05     $ &              $\leq 0.51     $ &                         $\leq0.64     $ &                            $6.2$ & $\leq -5.09$ &                                       \\[0.1cm]
 \multirow{2}{*}{XMM-Newton} &                      G 202-66 &                    $7.999 \pm 0.3$ &            $35.92 \pm 1.35$ &                       $31.43 \pm 1.18$ &                           $9.88$ & -3.44 &      \multirow{2}{*}{$0.81 \pm 0.08$} \\
                             &                      G 202-67 &                  $0.514 \pm 0.093$ &             $2.31 \pm 0.42$ &                        $4.84 \pm 0.88$ &                           $2.96$ & -4.11 &                                       \\[0.1cm]
 \multirow{2}{*}{XMM-Newton} &                      LTT 6326 &                  $0.084 \pm 0.027$ &             $0.39 \pm 0.12$ &                        $0.26 \pm 0.08$ &                          $15.39$ & -5.60 &        \multirow{2}{*}{$\geq 0.1     $} \\
                             &                      LTT 6325 &                     $\leq 0.08     $ &              $\leq 0.37     $ &                         $\leq 0.21     $ &                          $20.61$ & $\leq -5.74$ &                                       \\[0.1cm]
\multirow{2}{*}{Chandra/ACIS} &                     LP 320-163 &                  $0.098 \pm 0.109$ &             $0.25 \pm 0.28$ &                         $1.25 \pm 1.4$ &                           $0.91$ & -4.57 &   \multirow{2}{*}{/}    \\
                          &                     LP 320-162 &                    $\leq 0.564     $ &              $\leq1.48     $ &                         $\leq7.55     $ &                            $0.9$ & $\leq -3.78$ &                                       \\[0.1cm]
 \multirow{2}{*}{Chandra/ACIS} &                    LP 920-61 B &                  $6.634 \pm 2.549$ &             $2.69 \pm 1.03$ &                        $5.87 \pm 2.25$ &                            $2.8$ & -4.02 &        \multirow{2}{*}{$\geq 0.1     $} \\
                          &                    LP 920-61 A &                    $\leq 5.474     $ &              $\leq 2.23     $ &                          $\leq 4.7     $ &                           $2.93$ & $\leq -4.12$ &                                       \\[0.1cm]
 \multirow{2}{*}{Chandra/ACIS} &              SCR J0602-3952-B &                    $\leq 0.255     $ &              $\leq 0.65     $ &                         $\leq 1.25     $ &                           $3.32$ & $\leq -4.71$ &   \multirow{2}{*}{/}    \\
                          &              SCR J0602-3952-A &                    $\leq 0.342     $ &              $\leq0.88     $ &                         $\leq1.15     $ &                           $5.92$ & $\leq -4.83$ &                                       \\[0.1cm]
 \multirow{2}{*}{Chandra/ACIS} &                      Ross 868 &                 $142.48 \pm 6.377$ &             $20.09 \pm 0.9$ &                       $23.86 \pm 1.07$ &                           $6.65$ & -3.52 &      \multirow{2}{*}{$0.15 \pm 0.03$} \\
                          &                      Ross 867 &                $105.183 \pm 6.508$ &            $14.84 \pm 0.92$ &                        $33.97 \pm 2.1$ &                           $2.62$ & -3.25 &                                       \\[0.1cm]
 \multirow{2}{*}{Chandra/ACIS} &                  TIC 20446899 &                  $0.018 \pm 0.007$ &             $0.04 \pm 0.01$ &                        $0.02 \pm 0.01$ &                          $22.28$ & -6.79 &   \multirow{2}{*}{/}    \\
                          &  Gaia DR2 3532611086293698560 &                    $\leq 0.076     $ &              $\leq0.15     $ &                          $\leq0.2     $ &                           $6.09$ & $\leq -5.60$ &                                       \\[0.1cm]
 \multirow{2}{*}{Chandra/HRC} &                        G236-1 &                  $56.11 \pm 2.026$ &             $9.61 \pm 0.35$ &                        $7.56 \pm 0.27$ &                          $11.91$ & -4.09 &      \multirow{2}{*}{$0.18 \pm 0.02$} \\
                          &                        G236-2 &                  $71.16 \pm 2.281$ &            $12.24 \pm 0.39$ &                       $11.56 \pm 0.37$ &                           $8.62$ & -3.85 &                                       \\[0.1cm]
 \multirow{2}{*}{Chandra/ACIS} &   Gaia DR2 715928515183511040 &                    $\leq 1.464     $ &              $\leq 2.32     $ &                         $\leq 6.31     $ &                           $2.07$ & $\leq -3.95$ &   \multirow{2}{*}{/}    \\
                          &                  TIC 16151129 &                    $\leq 1.076     $ &              $\leq 1.69     $ &                         $\leq 2.15     $ &                           $6.18$ & $\leq -4.56$ &                                       \\[0.1cm]
 \multirow{2}{*}{Chandra/ACIS} &                    Ross 110 B &                    $\leq 0.235     $ &              $\leq 0.13     $ &                         $\leq 0.34     $ &                           $2.21$ & $\leq -5.22$ &    \multirow{2}{*}{/}   \\
                          &                    Ross 110 A &                    $\leq 0.189     $ &              $\leq 0.11     $ &                         $\leq 0.24     $ &                           $2.72$ & $\leq -5.41$ &                                       \\[0.1cm]

\hline
\end{tabular}
\end{table*}
\end{landscape}

\onecolumn
\addtocounter{table}{-1}
\begin{landscape}
\begin{table*}
	\caption{continued}

\begin{tabular}{llccccccc}
\hline
\hline

                     mission &                     component & $F_x \times 10^{-14}$ $[erg/s/cm^2]$ & $L_x \times 10^{27}$ [erg/s] & $Fx_{surf} \times 10^5$ $[erg/s/cm^2]$ & $L_{bol} \times 10^{31}$ [erg/s] &   $\log R_x$ & $\left|\log \frac{Fx_{surfA}}{Fx_{surfB}}\right|$ \\
\hline

\multirow{2}{*}{Chandra/HRC} & GJ 65 A &  $268.0 \pm 5.6$ & $2.39 \pm 0.05$ & $14.4 \pm 0.3$ & 0.62 & -3.41 & \multirow{2}{*}{$0.14 \pm 0.02$}\\
                         & GJ 65 B & $351.5 \pm 7.9$ & $3.10 \pm 0.07$ & $20.1 \pm 0.4$ & 0.52 & -3.22 & \\[0.1cm]

 \multirow{2}{*}{eROSITA} &                 TIC 206617096 &                  $5.721 \pm 0.339$ &               $8.4 \pm 0.5$ &                        $23.55 \pm 1.4$ &                           $1.99$ & -3.38 &      \multirow{2}{*}{$0.22 \pm 0.03$} \\
                          &                 TIC 206617113 &                 $10.637 \pm 0.574$ &            $15.63 \pm 0.84$ &                       $39.09 \pm 2.11$ &                           $2.32$ & -3.17 &                                       \\[0.1cm]
 \multirow{2}{*}{eROSITA} &                       GJ 3148 &                $145.246 \pm 1.889$ &            $31.85 \pm 0.41$ &                       $41.72 \pm 0.54$ &                           $5.87$ & -3.27 &      \multirow{2}{*}{$0.25 \pm 0.01$} \\
                          &                       GJ 3149 &                 $51.654 \pm 1.088$ &            $11.37 \pm 0.24$ &                        $23.54 \pm 0.5$ &                           $3.01$ & -3.42 &                                       \\[0.1cm]
 \multirow{2}{*}{eROSITA} &              UCAC4 235-004550 &                 $87.329 \pm 0.798$ &           $166.37 \pm 1.52$ &                      $105.83 \pm 0.97$ &                          $16.43$ & -2.99 &      \multirow{2}{*}{$0.27 \pm 0.01$} \\
                          &              UCAC4 235-004546 &                 $35.253 \pm 0.602$ &            $67.27 \pm 1.15$ &                       $57.42 \pm 0.98$ &                          $10.39$ & -3.19 &                                       \\[0.1cm]
 \multirow{2}{*}{eROSITA} &  Gaia DR2 4899032116649119616 &                   $5.881 \pm 0.22$ &            $10.88 \pm 0.41$ &                       $33.08 \pm 1.24$ &                            $1.8$ & -3.22 &   \multirow{2}{*}{/}    \\
                          &  Gaia DR2 4899029951985608576 &                   $\leq 27.522     $ &             $\leq51.32     $ &                       $\leq315.58     $ &                            $0.7$ & $\leq -2.14$ &                                       \\[0.1cm]
 \multirow{2}{*}{eROSITA} &                 TIC 167422188 &                   $1.354 \pm 0.02$ &             $2.75 \pm 0.04$ &                        $1.81 \pm 0.03$ &                          $15.46$ & -4.75 &      \multirow{2}{*}{$0.09 \pm 0.03$} \\
                          &                 TIC 167417695 &                  $0.794 \pm 0.048$ &              $1.63 \pm 0.1$ &                        $1.46 \pm 0.09$ &                           $9.41$ & -4.76 &                                       \\[0.1cm]
 \multirow{2}{*}{eROSITA} &                 TIC 106344480 &                $225.667 \pm 1.148$ &           $163.76 \pm 0.83$ &                      $100.37 \pm 0.51$ &                          $17.47$ & -3.03 &      \multirow{2}{*}{$0.07 \pm 0.06$} \\
                          &                 TIC 106493402 &                $84.167 \pm 10.925$ &            $61.45 \pm 7.98$ &                       $85.2 \pm 11.06$ &                           $5.37$ & -2.94 &                                       \\[0.1cm]
 \multirow{2}{*}{eROSITA} &                 TIC 416857959 &                  $21.059 \pm 0.45$ &            $39.32 \pm 0.84$ &                       $33.52 \pm 0.72$ &                          $10.42$ & -3.42 &       \multirow{2}{*}{$\geq 0.63     $} \\
                          &                 TIC 450297524 &                    $\leq 5.536     $ &             $\leq 10.34     $ &                         $\leq 7.89     $ &                          $12.44$ & $\leq -4.08$ &                                       \\[0.1cm]
 \multirow{2}{*}{eROSITA} &                  TIC 36765037 &                 $15.775 \pm 1.173$ &             $17.49 \pm 1.3$ &                       $37.93 \pm 2.82$ &                           $2.82$ & -3.21 &        \multirow{2}{*}{$\geq 0.4     $} \\
                          &                  TIC 36765044 &                    $\leq 5.905     $ &              $\leq 7.22     $ &                        $\leq 15.08     $ &                           $2.97$ & $\leq -3.61$ &                                       \\[0.1cm]
 \multirow{2}{*}{eROSITA} &                 TIC 151639642 &                   $4.742 \pm 0.24$ &              $7.83 \pm 0.4$ &                        $6.19 \pm 0.31$ &                          $11.84$ & -4.18 &      \multirow{2}{*}{$0.69 \pm 0.04$} \\
                          &              UCAC4 270-056947 &                 $16.486 \pm 1.323$ &            $27.29 \pm 2.19$ &                       $30.48 \pm 2.45$ &                           $7.11$ & -3.42 &                                       \\[0.1cm]
 \multirow{2}{*}{eROSITA} &              UCAC4 150-081944 &                  $7.455 \pm 0.727$ &             $8.39 \pm 0.82$ &                       $26.14 \pm 2.55$ &                           $1.75$ & -3.32 &      \multirow{2}{*}{$0.49 \pm 0.04$} \\
                          &                 TIC 317385747 &                 $62.761 \pm 0.868$ &            $70.36 \pm 0.97$ &                       $81.42 \pm 1.13$ &                           $6.84$ & -2.99 &                                       \\[0.1cm]
 \multirow{2}{*}{eROSITA} &                 TIC 392786054 &                  $10.25 \pm 0.548$ &             $8.68 \pm 0.46$ &                       $13.53 \pm 0.72$ &                           $4.49$ & -3.71 &      \multirow{2}{*}{$0.41 \pm 0.03$} \\
                          &                 TIC 392785968 &                  $16.011 \pm 0.35$ &             $13.66 \pm 0.3$ &                        $34.9 \pm 0.76$ &                           $2.25$ & -3.22 &                                       \\[0.1cm]
 \multirow{2}{*}{eROSITA} &                       WT 2090 &                 $24.581 \pm 0.953$ &            $13.76 \pm 0.53$ &                       $39.54 \pm 1.53$ &                           $1.93$ & -3.15 &       \multirow{2}{*}{$\geq 0.53     $} \\
                          &                     Wolf 1501 &                   $\leq 13.757     $ &              $\leq 7.69     $ &                        $\leq 11.58     $ &                           $4.73$ & $\leq -3.79$ &                                       \\[0.1cm]
 \multirow{2}{*}{eROSITA} &                 TIC 410458113 &                 $17.912 \pm 0.407$ &            $16.59 \pm 0.38$ &                       $14.99 \pm 0.34$ &                           $9.27$ & -3.75 &       \multirow{2}{*}{$\geq 0.46     $} \\
                          &              UCAC4 385-070621 &                    $\leq 3.753     $ &              $\leq 3.48     $ &                         $\leq 5.22     $ &                           $4.75$ & $\leq -4.14$ &                                       \\[0.1cm]
 \multirow{2}{*}{eROSITA} &                 TIC 229807000 &                $130.399 \pm 1.647$ &           $336.38 \pm 4.25$ &                       $206.17 \pm 2.6$ &                          $17.47$ & -2.72 &      \multirow{2}{*}{$0.28 \pm 0.01$} \\
                          &                 TIC 229807051 &                 $26.966 \pm 0.832$ &            $69.26 \pm 2.14$ &                      $109.43 \pm 3.38$ &                            $4.4$ & -2.80 &                                       \\
\hline
\end{tabular}
\end{table*}

\end{landscape}
\twocolumn

\subsection{Intrabinary X-ray surface flux difference}

\noindent Having estimated the X-ray surface flux of NLTT~41135 and NLTT~41136, and of the binaries in the control sample, as a next step, we estimated the coronal activity level difference between the two stars of each binary. The activity difference is calculated as the absolute logarithmic value of the ratio of the stellar X-ray surface fluxes of the two stars. In Table \ref{tab:fluxes}, the values of the stellar X-ray surface flux together with the activity difference ratio for each binary are given. The mean coronal activity difference in the X-ray regime between coeval M dwarfs in the control sample is $|\log(F\mathrm{x_{surfA}}/F\mathrm{x_{surfB}})| = 0.34 \pm 0.22$, with A and B denoting the surface flux of the primary and the secondary component, respectively. This means that, on average, the X-ray surface flux between two coeval M dwarfs can differ by a factor of $2.2 \pm 1.7$, which is similar to the intrinsic variability seen in single M dwarf stars in time-averaged X-ray data \citep{Marino2000,Stelzer2013,Magaudda2022}. The mean activity difference was calculated including systems where one star is undetected, and for these systems, we consider the activity difference given in Table \ref{tab:fluxes}. Since the sample has three systems with components separated by less than 100~AU, which is the lower limit for binary systems to be considered as wide \citep{Desidera2007}, we tested if a correlation between the activity difference indicator and the spatial separation between coeval stars exists, and found none.

We did not include the following systems in the analysis: TIC 293303829 / TIC 293303832, Gaia DR2 4899032116649119616 / Gaia DR2 4899029951985608576, LP 320-163 / LP 320-162, and TIC 20446899 / Gaia DR2 3532611086293698560. All systems have one undetected stellar component that has an upper limit value higher than the flux of the detected component, due to differences in the exposure times and in the detectors used. This means that the activity difference of each pair is unconstrained. Additionally, we did not include the undetected systems SCR J0602-3952 A/B, Ross~110~A/B, and Gaia~DR2~715928515183511040/TIC 16151129, where the activity difference between the components is unconstrained as well.

In Figure \ref{fig:gaia_comp}, the coronal activity difference is shown as a function of the absolute value of the $G-G_{\mathrm{RP}}$ color difference between the stars in each binary. The 68.3\% confidence interval of the control sample distribution is marked with the shaded region, and the mean of the distribution is presented as black solid line.
The coronal activity level difference between NLTT~41135 and NLTT~41136 is $|\log(F\mathrm{x_{surfA}}/F\mathrm{x_{surfB}})| = 1.09 \pm 0.18$ and is given as red asterisk symbol. Here, the brown dwarf-hosting M dwarf has an X-ray surface flux more than an order of magnitude higher than that of its stellar companion. Considering the X-ray surface flux of NLTT~41135 calculated including the flare emission, the activity difference rises to $|\log(F\mathrm{x_{surfA}}/F\mathrm{x_{surfB}})| = 1.16 \pm 0.18$.

\section{Discussion}

\subsection{The activity difference distribution}
\label{sample_discussion}
\noindent It is well known that stars of similar mass, but different ages have X-ray emission levels that can differ by several orders of magnitude (see e.g. \citealt{Guedel2004} and references therein). When, on the other hand, we consider stars of similar mass and age, as in our control sample, their emission and activity levels should be more consistent. As we have shown, two stars of the same spectral type in a binary will have a certain degree of difference in their activity level, but typically within a factor of two in coronal brightness. This is much smaller than the range observed for differently-aged stars.

This is also shown in Fig. \ref{fig:control_sample}, where the distribution of all investigated stars is shown in the X-ray surface flux -- {\it Gaia} color parameter space. There, individual stars are shown as black dots and those belonging to the same system are connected with a grey line. If both stars are detected, the connecting line is solid, otherwise, the line is dashed, which indicates that the slope of the connecting line is a minimum absolute value. The NLTT system is presented with red asterisks and a red connecting line; here we used the quiescent X-ray surface flux for the BD-hosting star.

The distribution of systems in Fig. \ref{fig:control_sample} shows a spread of more than three orders of magnitude in the X-ray surface fluxes. The stars in our sample, thus, span nearly the full range
of X-ray activity levels present in M dwarfs which was recently shown by \cite{Caramazza2023}
on a volume-complete sample to range from $<~10^4~\mathrm{erg/s/cm^2}$ (corresponding to the X-ray 
emission of solar coronal holes) to $>~10^7~\mathrm{erg/s/cm^2}$ (corresponding to solar cores of active regions and flares). The X-ray activity level of coronally active stars is known to be linked 
to the stellar rotation rate, which evolves over time. Therefore, our stars likely represent 
a range of ages. In fact, within a given binary the X-ray surface fluxes of the two components 
(which can be assumed to be coeval) are similar to each other, and this leads to the low
value of the average activity difference in our sample shown in Fig. \ref{fig:gaia_comp}. However, the activity levels of coeval stars are not equal to each other and there are several processes that can be considered as drivers of the activity difference distribution seen in Fig. \ref{fig:gaia_comp} and \ref{fig:control_sample}. In the following, we discuss phenomena like saturated coronal emission, the fully convective boundary, and activity cycles.

\subsubsection{Coronal saturation regime}

\noindent According to various studies, the coronae of cool main-sequence stars can operate in the saturated or unsaturated regime of emission \citep{Pallavicini1981,Pizzolato2003, Wright2011,Wright2018,Magaudda2020,Reiners2022,Magaudda2022}. In the first regime, the X-ray emission reaches a maximum value and does not depend on the stellar rotation rate, while in the latter, the X-ray emission decreases with the increasing rotation period of the star.

To estimate the coronal emission state, we calculated the parameter $\log R_{\mathrm{x}} = \log \frac{L_\mathrm{x}}{L_{\mathrm{bol}}}$ for each individual star in our sample. Here, we calculated the bolometric luminosity, similar to the stellar radius in Sec. \ref{sec:radius}, by using the $L_{\mathrm{bol}}$ values for main-sequence stars published by \cite{Pecaut2013}, and interpolating them over the absolute {\it G} magnitudes given in Table \ref{tab:photo_radii}. The $\log R_\mathrm{x}$ values of individual stars are given in Table \ref{tab:fluxes}.

We estimated the saturation limit for M dwarf stars to be $\log R_\mathrm{x}~=~-3.26^{+0.38}_{-0.36}$, by employing the saturation values for different mass bins from Fig. 9b by \cite{Magaudda2022}, and estimating the average value for their full mass range. Taking the average lower saturation limit of $\log R_\mathrm{x} = -3.62$, we find that 18 of the pairs in our control sample have at least one star in the saturated regime. These {\it high-activity} systems are also seen in Fig. \ref{fig:control_sample} with X-ray surface fluxes above $F\mathrm{x_{surf}} \approx 10^6 \mathrm{erg/s/cm^2}$. Having the majority of the control systems in the high-activity regime is most probably due to the fact that we employ archival data, where the probability of detecting active stars is higher than for low-activity stars. Therefore, our control sample is biased toward brighter X-ray stars.

One might argue that the high-activity systems are more likely to have equal surface fluxes for both stars because they are saturated, which is often interpreted to mean that most of their corona is full with X-ray-emitting magnetic structures. Consequently, it is difficult (or impossible) to produce a coronal activity difference of one order of magnitude between two coeval saturated stars. However, the control sample also includes seven low-activity systems (where both M~dwarfs have $F\mathrm{x_{surf}} \lesssim 10^6 \mathrm{erg/s/cm^2}$). Unsaturated stars are not expected to have a coronal filling factor near 100\%, and thus the coronal activity difference between the two components of such binaries may take larger values than found in highly active, saturated systems. However, we measure for the low-activity stars $|\log(F\mathrm{x_{surfA}}/F\mathrm{x_{surfB})| = 0.30}$, lower than the average for all systems. Here, it has to be noted that some low-activity systems have one star undetected and the calculated mean activity difference is a lower limit.

\subsubsection{Fully convective boundary}

\noindent The fully convective boundary occurs in mid-M dwarfs (e.g. \citealt{Copeland1970,Chabrier1997}) and is encompassed by the sample selection we made. It is assumed that, due to high opacity in later M dwarfs, efficient energy transport inside the star is possible only through convection and the star becomes fully convective. 

While it was expected that the coronal properties might change at this boundary due to a switch in dynamo mode from an $\alpha-\Omega$ dynamo to a different one (e.g. \citealt{Chabrier2006}), observations showed that there is no abrupt change in X-ray luminosity or other coronal activity indicators \citep{Stelzer2013, Wright2016}. Rather, changes in coronal properties were found at the very low-mass end of the M dwarf sequence, where the low temperature of the photosphere may start to affect the formation of active regions \citep{Berger2010, Robrade2009, Stelzer2012, Williams2014}.

In our sample, we have a few pairs that straddle the fully convective boundary. If this boundary is set at the spectral type $\approx$ M4 with the {\it Gaia} color of $G-G_{\mathrm{RP}} \approx 1.24$, in eight systems the primary is partially convective while the secondary is fully convective. However, in these systems, the less-massive star has the spectral type M4 and not later. Therefore, we cannot with certainty say that it is a fully convective star, only that it is potentially fully convective. Therefore, any potential effect due to having fully convective stars in our sample might not be present in our sample.

\subsubsection{Activity cycles}

\noindent One aspect of magnetic activity we were not able to account for is the activity cycle of stars in the X-ray regime. It is well established that stars other than the Sun can have these kinds of cycles \citep{Hempelmann2006,Robrade2007,Favata2008,DeWarf2010,Coffaro2020}; however, activity cycles in saturated stars and stars close to the fully convective boundary seem to be elusive not only in the coronal part of the stellar atmosphere but in the chromosphere as well \citep{Robertson2013,Fuhrmeister2023}. It is hypothesized that activity cycles in fully convective stars are absent because the efficiency of magnetic braking decreases (e.g. \citealt{Chabrier2006}). For saturated stars, activity cycles might be absent because their coronae are fully covered with magnetic X-ray emitting structures leaving no space for additional X-ray emitting regions. In fact, none of the stars with a detected activity cycle is saturated (Drake \& Stelzer, in prep.). However, since in the control sample, we have stars that are both unsaturated and partly convective, the existence of X-ray activity cycles cannot be fully excluded, and their contribution to the observed activity difference remains unconstrained.

\subsection{The coronal activity level difference between NLTT~41135 and NLTT~41136 and its physical interpretation}

\noindent The spin evolution of a star with a close-in companion is driven by processes that can have opposite effects: magnetic braking and tidal interactions. While the wind-driven braking slows down the stellar rotation rate, tidal interactions with a close-in companion can induce spin-up via angular momentum transfer if the configuration of the system is such that the orbital rate of the companion is greater than the rotation rate of the star. As a consequence of the spin-up scenario, the star can experience a higher magnetic activity level than would be the case without tidal interactions. To test this hypothesis when the companion is a substellar object, we analysed the subsystem NLTT~41135, a brown dwarf-M dwarf pair, together with their stellar companion NLTT~41136, which is expected to have the baseline activity level which is governed only by magnetic braking. 

Since, in general, a difference in activity between coeval stars with the same spectral type should be expected, we introduced a control sample which showed that the average difference in activity between these stars is $|\log(F\mathrm{x_{surfA}}/F\mathrm{x_{surfB}})| = 0.34 \pm 0.22$. In comparison to that sample, the coronal activity level difference in the NLTT~41135/41136 system is $|\log(F\mathrm{x_{surfA}}/F\mathrm{x_{surfB}})| = 1.1 \pm 0.2$. This result makes the difference in the coronal activity level between the two stars of this system highly significant -- it is at a $\approx 3.44\sigma$ level of the control sample. This value is, however, an upper limit since the control sample also consists of systems with one undetected component. If we consider only systems with both stars detected, the significance level of the activity difference in the NLTT~41135/41136 system rises to $3.7\sigma$, where the mean activity difference of the detected binaries is $|\log(F\mathrm{x_{surfA}}/F\mathrm{x_{surfB}})| = 0.31 \pm 0.21$.

The possible sources of activity difference between the stellar components of binary systems in our control sample were discussed in Section \ref{sample_discussion}. Considering their impact on the activity difference in the NLTT~41135/41136 system, the effect of the fully convective boundary should be negligible since both stars are, given their {\it Gaia} color, fully convective, as well as the effect of the coronal emission regime since both stars have unsaturated X-ray emission. The effects we cannot quantify are the possible impact of activity cycles and the stochastic, short-term X-ray variability. However, the control sample provides a good estimate of the combined effect of all aforementioned phenomena and confirms the high significance of tidal interactions occurring between the brown dwarf and its host.

Furthermore, we can also exclude that the brown dwarf itself contributes significantly to the observed X-ray photons. Only very young brown dwarfs at ages of a few million years have been found to be X-ray emitters \citep{Neuhaeuser1999,Mokler2002,Preibisch2005}, with flaring events providing their peak luminosity \citep{Rutledge2000,Stelzer2004}. Old brown dwarfs are found to be X-ray quiet with $\log L\mathrm{_x [erg/s]} \lesssim 25$ \citep{Stelzer2006}. Therefore, with the kinematic age of our target system being at least 1 Gyr \citep{Irwin2010}, the quiescent X-ray emission from the brown dwarf corona can be considered insignificant.

We therefore physically interpret the activity difference as a consequence of star-brown dwarf interaction. Tidal interactions between a slowly rotating star and its quickly orbiting satellite are expected to lead to a transfer of angular momentum from the orbit of the satellite into the spin of the star \citep{Zahn1977}. We do not have direct information about the rotational period of NLTT~41135 in order to compare it to the rotation of its stellar companion. However, we can estimate their expected rotational periods from activity-rotation relationships. We use the relationships from \citet{Wright2018} for fully convective stars, respectively, to estimate the convective turnover times and then, from the activity indicator $R_\mathrm{x}$, the stellar rotation period. We find an expected rotation period of ca.\ 36 days for NLTT~41135, and a much longer expected rotation period for NLTT~41136 of the order of 97 days. If NLTT~41135 had a similarly low activity level as its stellar companion, i.e.\ $\log R_\mathrm{x} \approx -4.9$, we would expect a rotation period of roughly 114 days.

Keeping in mind that these expected rotation periods have large uncertainties, we perform an order-of-magnitude estimate of the angular momentum transfer that we expect to have taken place to make NLTT~41135 rotate at a ca.\ 36-day period instead of a 114-day period. If we approximate the angular momentum of NLTT~41135 with that of a rotating solid sphere ($L_{\mathrm{ang}}=2/5 M_\ast R_\ast \omega$, with $\omega = 2\pi/P_{\mathrm{rot}}$ being its rotational frequency, and $M_{\ast}$ and $R_{\ast}$ being the stellar mass and radius, respectively), the difference in angular momentum between the 36-day and the 114-day rotational states amounts to about $2.5\times 10^{36}~\mathrm{\,g\,cm/s}$. The present-day orbital motion of the brown dwarf has an angular momentum of $L_{\mathrm{orb}} = a_{\mathrm{sem}} M_{\mathrm{BD}} v_{\mathrm{BD}} \sim 1.7 \times 10^{50}$~$\mathrm{\,g\,cm/s}$, with $M_{\mathrm{BD}}$ and $v_{\mathrm{BD}}$ being the mass and the orbital velocity of the brown dwarf, respectively, and $a_{\mathrm{sem}}$ being the orbital semi-major axis. The brown dwarf's orbital angular momentum is more than 10 orders of magnitude larger than the star's rotational angular momentum. Therefore, even a slight shrinking of the brown dwarf's orbit can easily supply enough angular momentum to spin up the central star to the observed levels. We, therefore, conclude that tidal interactions between low-mass stars and brown dwarfs is indeed a viable scenario for stellar spin-up.



\subsection{The difference in observed energy ranges as a source of activity difference}

\noindent One technical aspect that has to be considered in the discussion of the origin of activity differences in all our systems is the different energy ranges that are encompassed by the various instruments we use. As we discuss in Appendix \ref{appA} and show in Table \ref{tab:app}, the eROSITA and {\it Chandra}/HRC instruments - with the energy bands of 0.2-2.3 and 0.08-10.0 keV, respectively - are collecting a similar amount of energy as if they were to observe within the canonical energy band of 0.2-2.0 keV for the given coronal emission and temperature\footnote{The stellar coronae of cool stars emit the bulk of their magnetically induced high energy radiation in the 0.2-2.0 keV band (see e.g. \citealt{Gudel1997}).}. The differences in fluxes that arise from the difference in the observed energy bands are well within the coronal activity difference uncertainty given in Table~\ref{tab:fluxes}. Also, we are interested in the flux ratio of two stars observed in the same energy band; therefore, the mismatch in energy bands should not affect the reliability of our results.

\section{Summary and conclusion}

\noindent To estimate the significance of tidal interactions between a star and its close-in companion, we analysed the X-ray observation of the system NLTT~41135/41136 taken by the {\it Chandra} X-ray Observatory. Here, the NLTT~41135 component consists of an M5V dwarf and a T6-T8 brown dwarf in close orbit, while their M4V dwarf companion, NLTT~41136, is the primary star of the system. Previous radial velocity and astrometric measurements have indicated that the whole system is kinematically old and belongs to the thick Galactic disk \citep{Irwin2010}. Our observations show that the quiescent X-ray surface flux of NLTT~41135 is more than an order of magnitude higher than that of NLTT~41136.

To put this flux difference in context, we calculated the X-ray surface fluxes of stars in 25 wide binary systems consisting of M dwarf stars similar in stellar parameters. We found the mean activity difference in these systems - the activity difference parameter being the absolute value of the logarithm of the surface flux ratio - to be $|\log(F\mathrm{x_{surfA}}/F\mathrm{x_{surfB}})| = 0.34 \pm 0.22$, while the same parameter for the NLTT~41135/41136 system has the value of $|\log(F\mathrm{x_{surfA}}/F\mathrm{x_{surfB}})| = 1.1 \pm 0.2$. This result makes the BD-hosting system a $3.44\sigma$ outlier.

We found that in some of our reference systems, stars were in different coronal emission regimes, were likely on different sides of the boundary between partially and fully convective M dwarfs, and showed short-term stochastic variability. On the other hand, NLTT~41135 and NLTT~41136 are both fully convective, operate in the unsaturated emission regime, and have their quiescent activity level compared to one another. Therefore, the observed excess in the coronal activity of the brown dwarf-host NLTT~41135 is most likely induced by the spin-up process due to angular momentum transfer from the brown-dwarf orbit to the stellar spin via tidal interactions.

This is the first study that quantifies the impact of close-in brown dwarfs on the evolutionary path of main-sequence, low-mass stars. The estimated change in rotation period of $\approx 80$ days and measured increase in the coronal activity level by one order of magnitude question the reliability of these parameters as proxies for the stellar age of main-sequence stars. This being said, tidal interactions might not be the only type of interaction occurring in these types of systems. Although tidal interactions - due to the mass ratio of $\approx$~5:1 - most likely play a significant role between NLTT~41135 and its orbiting brown dwarf, as in a binary system consisting of a solar-type star and an M dwarf star, the existence of magnetic interactions between these two objects and their impact on NLTT~41135 remains an open question. Therefore, more studies of wide binary systems with and without close-in companions will improve our understanding of the impact the different types of interactions might have on the evolution of low-mass, main-sequence stars.

\section*{Acknowledgements}

\noindent The authors thank the anonymous referee for their valuable comments. The authors thank Dus{\'a}n Tubin for providing eROSITA-derived upper limits on the X-ray fluxes for several stars, Dr. Enza Magaudda for providing saturation limits for M dwarf stars in the saturation regime, and Dr. Matthias Mallonn for a fruitful discussion on the control sample for the presented analysis.
NI and KP acknowledge support from the German Leibniz-Gemeinschaft under project number P67/2018. M.A.A.~acknowledges support from a Fulbright U.S.~Scholar grant co-funded by the Nouvelle-Aquitaine Regional Council and the Franco-American Fulbright Commission. M.A.A.~also acknowledges support from a Chr\'etien International Research Grant from the American Astronomical Society.
The scientific results reported in this article are based in part on observations made by the {\it Chandra} X-ray Observatory.
This research has made use of data obtained from the 4XMM XMM-Newton serendipitous source catalogue compiled by the 10 institutes of the XMM-Newton Survey Science Centre selected by ESA.
This work is based on data from eROSITA, the soft X-ray instrument aboard SRG, a joint Russian-German science mission supported by the Russian Space Agency (Roskosmos), in the interests of the Russian Academy of Sciences represented by its Space Research Institute (IKI), and the Deutsches Zentrum für Luft- und Raumfahrt (DLR). The SRG spacecraft was built by Lavochkin Association (NPOL) and its subcontractors, and is operated by NPOL with support from the Max Planck Institute for Extraterrestrial Physics (MPE). The development and construction of the eROSITA X-ray instrument was led by MPE, with contributions from the Dr.\ Karl Remeis Observatory Bamberg \& ECAP (FAU Erlangen-N\"urnberg), the University of Hamburg Observatory, the Leibniz Institute for Astrophysics Potsdam (AIP), and the Institute for Astronomy and Astrophysics of the University of Tübingen, with the support of DLR and the Max Planck Society. The Argelander Institute for Astronomy of the University of Bonn and the Ludwig Maximilians Universit\"at M\"unchen also participated in the science preparation for eROSITA. The eROSITA data shown here were processed using the eSASS software system developed by the German eROSITA consortium. 
This research has made use of the VizieR catalogue access tool, CDS,
Strasbourg, France (DOI : 10.26093/cds/vizier). The original description 
of the VizieR service was published in 2000, A\&AS 143, 23.
This research made use of Astropy,\footnote{\url{http://www.astropy.org}} a community-developed core Python package for Astronomy \citep{Astropy2013, Astropy2018}. 
This work has made use of data from the European Space Agency (ESA) mission
{\it Gaia}\footnote{\url{https://www.cosmos.esa.int/gaia}}, processed by the {\it Gaia}
Data Processing and Analysis Consortium (DPAC\footnote{\url{https://www.cosmos.esa.int/web/gaia/dpac/consortium}}). Funding for the DPAC has been provided by national institutions, in particular the institutions
participating in the {\it Gaia} Multilateral Agreement.

\section*{Data Availability}

\noindent The eROSITA data used in this work will be part of an international data release and will be accessible via the eROSITA-DE Science Portal\footnote{\url{https://erosita.mpe.mpg.de/}}.
The {\it Chandra} and {\it XMM-Newton} data used in this work are publicly available at the {\it Chandra} Data Archive\footnote{\url{https://cda.harvard.edu/chaser/}} and the {\it XMM~-~Newton} Science Archive\footnote{\url{https://nxsa.esac.esa.int/nxsa-web/\#search}}. The source ID of each observation used in this research is given in Appendix \ref{AppB}.



\bibliographystyle{mnras}
\bibliography{references} 




\appendix

\section{Energy range flux comparison}
\label{appA}

\begin{table*}
	\centering
	\caption{Given are the input energy bands in which systems in our sample were observed, an assumed input flux $F_{\mathrm{xi}}$, and the corresponding output flux $F_{\mathrm{xo}}$ in the canonical energy band for various coronal temperatures. The parameter $\mathrm{|\log(F_{\mathrm{xi}}/F_{\mathrm{xo}})|}$ shows how much the activity difference parameter is affected due to the difference in energy bands.}
	\label{tab:app}
	\begin{tabular}{ccccccccccccccc} 
		\hline
		input energy band [keV] & output energy band [keV] & $\log_{10} T$ [K] & $F_{\mathrm{xi}}$ [$\mathrm{erg/s/cm^{2}}$] & $F_{\mathrm{xo}}$ [$\mathrm{erg/s/cm^{2}}$] & $\mathrm{|\log(F_{\mathrm{xi}}/F_{\mathrm{xo}})|}$ \\
        \hline
        \multirow{3}{*}{0.08 - 10.0} & \multirow{3}{*}{0.2 - 2.0} & 6.0 & \multirow{3}{*}{$10^{-14}$} & $10^{-14}$ & 0.0\\
        &  & 6.5 &  & $9.992 \times 10^{-15}$  & 0.00035\\
        &  & 7.0 &  & $9.453 \times 10^{-15}$ & 0.024\\
        \hline
        \multirow{3}{*}{0.2 - 2.3} & \multirow{3}{*}{0.2 - 2.0} & 6.0 & \multirow{3}{*}{$10^{-14}$} & $10^{-14}$ & 0.0\\
        &  & 6.5 &  & $9.995 \times 10^{-15}$  & 0.00022\\
        &  & 7.0 &  & $9.811 \times 10^{-15}$  & 0.008\\
        \hline
        \hline
	\end{tabular}
\end{table*}

\noindent The M dwarf binary systems considered here have been observed with various X-ray instrument set-ups and cover different energy ranges. From our control sample, eROSITA has observed 14 systems, nine systems were observed with {\it XMM-Newton's EPIC} camera, and eight systems with {\it Chandra}, out of which two were observed with HRC-I and HRC-S, and the rest with the ACIS instrument. NLTT~41135/41136 was observed with HRC-I. The eROSITA systems are observed in the 0.2-2.3 keV energy range; the HRC observations, both with the imager (I) and the spectrometer (S), are made in the 0.08-10.0 keV range, while the {\it XMM-Newton} systems have their flux estimated in the 0.2-2.0 keV range.

In Table \ref{tab:app}, we show how the flux observed in one of the considered bands converts to the flux in the canonical 0.2-2.0 keV band. For this task, we used the online tool {\sc webpimms} (v4.11a). As the parameter that shows the flux difference due to different energy bands, we used the absolute value of the logarithm of the flux ratio. This is a good representation of the impact the difference in the energy band will have on the activity difference parameter we use since it is also represented by the absolute value of the logarithm of the surface flux ratio. Although, we calculate the X-ray surface flux of stars, which aside from the observed flux, needs the knowledge of stellar radius and distance, not considering these values here is appropriate since we compare the flux of the same star in different energy bands.

\section{Notes on individual systems}
\label{AppB}

Here, we provide analysis details on systems that were observed with the XMM-Newton Space Observatory and the Chandra X-ray telescope. In general, the extraction region for sources observed with {\it XMM-Newton} has a radius of $15\arcsec$, while the radius of the background extraction region is $60\arcsec$. For {\it Chandra} sources, the extraction region has a radius of $1.5\arcsec$, while the background extraction region has a radius of $15\arcsec$. If the extraction radii differ from these values, we note it in the table.

We examined the X-ray light curve of each observation for flaring events and found one strong flare (peak count rate is $5-7 \times$ the quiescent count rate) in the light curve of GJ-15 A. The X-ray flux given in Table \ref{tab:fluxes} for this source is calculated excluding the time during which the flare occurred. All other sources observed with {\it XMM-Newton} and {\it Chandra} show occasional fluctuation, but no significant increase in count rate.

For stars where we could not determine the hardness ratio due to the detector properties or insufficient counts, we assumed the coronal temperature to be $\log_{10} T \mathrm{[K]} = 6.477$. The two bands used to estimate the hardness ratio are the soft band covering the range S~=~0.2-0.7 keV, and the hard band covering the range H~=~0.7-2.0 keV. The hardness ratio is estimated via HR~=~(H-S)/(H+S) (see \cite{Ilic2022} for details).

\begin{table*}
\caption{Notes on individual systems}
	\label{tab:notes}
\begin{tabular}{lp{1.5cm}cp{1cm}cccccc}

\hline
\hline
component & obs ID & mission & camera & HR & $\log_{10} T$ [K] & NOTES \\
\hline
Ross 868 & \multirow{2}{*}{\parbox{1.5cm}{500670201 \newline 500670301 \newline 500670401}} & \multirow{2}{*}{XMM} & \multirow{2}{*}{pn} & -0.171 & 6.593 & \multirow{2}{*}{$r_{source} = 12.0\arcsec$}\\[0.1cm]
Ross 867 & & & & -0.204 & 6.582 & \\[0.2cm]
GJ-15 A & \multirow{2}{*}{801400301} & \multirow{2}{*}{XMM} & \multirow{2}{*}{\parbox{1cm}{pn\newline MOS1 \newline MOS2}} & -0.468 & 6.460 & \\[0.1cm]
GJ-15 B & & & & -0.608 & 6.396 & \\[0.2cm]
TIC 355790951 & \multirow{2}{*}{406540301} & \multirow{2}{*}{XMM} & \multirow{2}{*}{\parbox{1cm}{pn\newline MOS2}} & -0.438 & 6.501 & \\[0.1cm]
TIC 355790950 & & & & / & 6.477 & \\[0.2cm]
Gaia DR2 1608710752684301312 & \multirow{2}{*}{804270201} & \multirow{2}{*}{XMM} & \multirow{2}{*}{\parbox{1cm}{pn\newline MOS2}} & -0.16 & 6.597 & \\[0.1cm]
Gaia DR2 1608710791338814208 & & & & -0.289 & 6.552 & \\[0.2cm]
TIC 436632332 & \multirow{2}{*}{743070301} & \multirow{2}{*}{XMM} & \multirow{2}{*}{\parbox{1cm}{pn\newline MOS1 \newline MOS2}} & -0.67 & 6.368 & \multirow{2}{*}{$r_{source} = 12.0\arcsec$}\\[0.1cm]
TIC 436632331 & & & & -0.213 & 6.556 & \\[0.2cm]
TIC 293303829 & \multirow{2}{*}{211280101} & \multirow{2}{*}{XMM} & \parbox{1cm}{MOS1 \newline MOS2} & / & 6.477 & \\[0.2cm]
TIC 293303832 & & & pn & / & 6.477 & \\[0.2cm]
LTT 6326 & \multirow{2}{*}{550970101} & \multirow{2}{*}{XMM} & \multirow{2}{*}{\parbox{1cm}{pn\newline MOS1 \newline MOS2}} & / & 6.477 &  \multirow{2}{*}{$r_{source} = 13.0\arcsec$}\\[0.1cm]
LTT 6325 & & & & / & 6.477 & \\[0.2cm]
G 202-66 & \multirow{2}{*}{605000501} & \multirow{2}{*}{XMM} & \multirow{2}{*}{\parbox{1cm}{pn\newline MOS2}} & -0.345 & 6.533 &  \multirow{2}{*}{$r_{source} = 12.0\arcsec$}\\[0.1cm]
G 202-67 & & & & / & 6.477 \\[0.2cm]
Gaia DR2 3074577322667614976 & \multirow{2}{*}{800400601} & \multirow{2}{*}{XMM} & \parbox{1cm}{pn\newline MOS1 \newline MOS2} & -0.167 & 6.594 & \\[0.3cm]
Gaia DR2 3074630580262065536 & & & {pn \newline MOS2} & -0.03 & 6.642 & \\[0.2cm]
Gaia DR2 4768120070358120064 & \multirow{2}{*}{744400301} & \multirow{2}{*}{XMM} &\multirow{2}{*}{pn} & / & 6.477 & \\[0.1cm]
Gaia DR2 4768120104717857792 & & & & / & 6.477 & \\[0.2cm]
LP320-163 & \multirow{2}{*}{5767} & \multirow{2}{*}{Chandra} & \multirow{2}{*}{ACIS-I} & / & 6.477 & \multirow{2}{*}{$r_{source} = 1\arcsec$}\\[0.1cm]
LP320-162 & & & & / & 6.477 & \\[0.2cm]
LP920-61 A & \multirow{2}{*}{\parbox{1cm}{13585 \newline 13588}} &  \multirow{2}{*}{Chandra} & \multirow{2}{*}{ACIS-I} & / & 6.477 & \\[0.1cm]
LP920-61 B &  &  & & / & 6.477 & \\[0.2cm]
SCR J0602-3952-A & \multirow{2}{*}{\parbox{1cm}{3202 \newline 3450}} & \multirow{2}{*}{Chandra} & \multirow{2}{*}{ACIS-I} & / & 6.477 & \multirow{2}{*}{$r_{source} = 1.2\arcsec$}\\[0.1cm]
SCR J0602-3952-B & & & & / & 6.477 & \\[0.2cm]
Ross 868 & \multirow{2}{*}{\parbox{1cm}{1453\newline 3224 \newline 4361}} & \multirow{2}{*}{Chandra} & \multirow{2}{*}{ACIS-I} & 0.695 & 6.695 & \\[0.1cm]
Ross 867 & & & & 0.587 & 6.641 & \\[0.2cm]
TIC 20446899 & \multirow{2}{*}{915} & \multirow{2}{*}{Chandra}  & \multirow{2}{*}{ACIS-S} & 0.585 & 7.0 & \\[0.1cm]
Gaia DR2 3532611086293698560 & & & & / & 6.477 & \\[0.2cm]
G236-1 & \multirow{2}{*}{6655} & \multirow{2}{*}{Chandra} & \multirow{2}{*}{HRC-I} & / & 6.7 & \multirow{2}{*}{$r_{source} = 1.2\arcsec$}\\[0.1cm]
G236-2 & & & & /& 6.7 & \\

\hline
\end{tabular}
\end{table*}

\addtocounter{table}{-1}
\begin{table*}
\caption{Notes on individual systems: continued}
\begin{tabular}{lp{1.5cm}cp{1cm}cccccc}

\hline
\hline
component & obs ID & mission & camera & HR & $\log_{10} T_{cor}$ [K] & NOTES \\
\hline
Gaia DR2 715928515183511040 & \multirow{2}{*}{16057} & \multirow{2}{*}{Chandra} & \multirow{2}{*}{ACIS-I} & / & 6.477 & \\[0.1cm]
TIC 16151129 & & & & / & 6.477 & \\[0.2cm]
Ross 110 A & \multirow{2}{*}{7607} & \multirow{2}{*}{Chandra} & \multirow{2}{*}{ACIS-I} & / & 6.477 & \\[0.1cm]
Ross 110 B & & & & /& 6.477 & \\

\hline
\end{tabular}
\end{table*}


\bsp	
\label{lastpage}
\end{document}